\documentclass[11pt]{article}
\pdfoutput=1

\usepackage{multirow}
\usepackage{amsmath, amsfonts, amssymb}
\usepackage{comment}
\usepackage{graphicx}
\usepackage{psfrag}
\usepackage{amsthm}
\usepackage[usenames,dvipsnames,svgnames,table]{xcolor}
\usepackage{enumerate}
\usepackage{tcolorbox}

\usepackage{mathtools}

\usepackage{soul}
\usepackage{subfig}
\usepackage{mathrsfs}  
 \usepackage{a4wide}
  \usepackage{tikz}
  \usepackage{tikz-cd}

  \usepackage{color}
  \definecolor{dark-gray}{gray}{0.20}
  \definecolor{gray}{gray}{0.30}
  \definecolor{light-gray}{gray}{0.80}
  \definecolor{dark-red}{rgb}{0.7,0,0}
  \definecolor{dark-green}{rgb}{0.1,0.4,0}
  \definecolor{dark-blue}{rgb}{0.3,0.3,0.7}
  \definecolor{light-blue}{rgb}{0.8,0.8,2}
      \definecolor{swamp}{RGB}{240, 199, 197}
     \definecolor{landscape}{RGB}{180, 250, 199}
          \definecolor{undecided}{RGB}{252, 252, 197}
  
  \usepackage{pifont}
\usepackage{setspace}
\usepackage{multirow}

\newcommand{\be}{\begin{equation}}
\newcommand{\ee}{\end{equation}}
\newcommand{\eq}[1]{(\ref{#1})}

\def\be{\begin{equation}}
\def\ee{\end{equation}}
\def\bea{\begin{eqnarray}}
\def\eea{\end{eqnarray}}

\def\simleq{\; \raise0.3ex\hbox{$<$\kern-0.75em
      \raise-1.1ex\hbox{$\sim$}}\; }
   \def\simgeq{\; \raise0.3ex\hbox{$>$\kern-0.75em
      \raise-1.1ex\hbox{$\sim$}}\; }

\numberwithin{equation}{section}

\usepackage{jheppub} 
\hypersetup{
    colorlinks=true,
    linkcolor=dark-blue,
    citecolor=dark-red,
    urlcolor=dark-green,
    linktoc=page,
    pageanchor=false
}
\title{\centering New Supersymmetric String Theories\\ from Discrete Theta Angles}

\author{Miguel Montero$^1$}
\author{ and H\'{e}ctor Parra De Freitas$^2$}

\affiliation{$^1$Jefferson Physical Laboratory, Harvard University,\\
Cambridge, MA 02138, USA}
\affiliation{$^2$Institut de Physique Th\'{e}orique, Universit\'{e} Paris Saclay, CEA, CNRS\\
Orme des Merisiers, 91191 Gif-sur-Yvette CEDEX, France.}

\emailAdd{mmontero@g.harvard.edu, hector.parradefreitas@ipht.fr}

\abstract{We describe three previously unnoticed components of the moduli space of minimally supersymmetric string theories in $d\geq 7$, describing in some detail their spectrum and duality properties. We find a new component in nine and eight dimensions, and two additional ones in seven dimensions.  These theories were originally discovered in a bottom-up classification of possible F/M-theory singularity  freezing patterns in the K3 lattice, described in a companion paper. The 9d/8d component can be understood as F/M-theory on a twisted fibration of the Klein bottle over a circle, while the new seven-dimensional components are described as IIB on Bieberbach manifolds with a duality bundle and RR-NSNS backgrounds turned on. All the new components  can be obtained from previously known theories by turning on certain discrete theta angles; however, the spectrum of massive objects is very different, and most strikingly, they feature an incomplete lattice of BPS strings, showing that string BPS completeness is not true in general even with sixteen supercharges. In all cases we find non-BPS representatives for each value of the charge, so the Completeness Principle is satisfied. We also analyze analogous theta angles in nonsupersymmetric string theories, and provide a detailed explanation of why the Type I discrete $\theta$ angle proposed in \href{https://arxiv.org/abs/1304.1551}{1304.1551} is unphysical, using this to clarify certain non-perturbative phenomena in $O8$ planes.}

\begin{document}

\makeatletter
\let\old@fpheader\@fpheader

\makeatother

\maketitle

\section{Introduction}\label{sec:intro}

There is much more to quantum gravity than what one can readily see at low energies. The spectrum of massive string states and their interactions are notoriously difficult to understand in most perturbative corners, and essentially untamed away from these. The spectrum of massive excitations is often related to the existence of \emph{discrete} theta angles: topological couplings that are controlled by a discrete parameter. A field theory example is given by $\mathcal{N}=4$ SYM with $SO(N)$ gauge group. This group has an ordinary theta angle associated to the first Pontryagin class of  $SO(N)$; but there is also a discrete coupling
\begin{equation*}\theta\int w_2^2\end{equation*}
to the second Stiefel-Whitney class of the $SO(N)$ bundle \cite{Aharony:2013hda}. As this class is $\mathbb{Z}_2$-valued, the parameter theta above can only take the values $0$ or $\pi$. The spectrum of extended operators and the massive excitations when the theory is put on an arbitrary manifold is sensitive to this theta angle.

Unlike their continuous counterparts, discrete theta angles in quantum gravity are not associated to moduli; therefore, they are invisible to the low energy supergravity. In other words, one can have two different quantum theories of gravity, with exactly the same low-energy lagrangian, which differ only by a discrete theta angle. The goal of this paper is to begin the exploration of discrete theta angles in quantum gravity in some generality.  As we will see, discrete theta angles lead to new supersymmetric string theories with sixteen supercharges, providing the first new models in this class since \cite{Hellerman:2005ja,Hellerman:2006tx}.

Explicit examples of quantum gravities with discrete theta angles were constructed by Cecotti and Vafa in \cite{Cecotti:2018ufg} via compactification in rigid Calabi-Yau threefolds. These models provide a stringy resolution of the strong CP problem where the theta angle is simply frozen to be either 0 or $\pi$. But in general, the physics of discrete theta angles is quite subtle (see \cite{Keurentjes:2001cp} for early work, involving six-dimensional compactifications with a $C_4$ discrete theta angle, which is very similar to what is presented in this paper). One outstanding example of these subtleties is the Sethi string \cite{Sethi:2013hra}, which amounts to turning on a certain RR discrete theta angle in the usual construction of type I string theory. While at first this seems like a consistent model, one runs into several paradoxes when compactifying and trying to track the meaning of theta across the duality web. This is a confusing state of affairs, with prior literature being unclear as to whether the Sethi string is part of the Landscape or the Swampland.

In this paper we resolve this question -- we figure out in which cases there is actually a theta angle and in which ones there isn't. We show, in particular, that the Sethi string is completely equivalent to the ordinary type I string\footnote{After completing this paper, we learned of a talk given by O. Bergman at Oviedo in 2015 \cite{Bergman2015} which essentially arrives at the same conclusion.}. But we also find the following theories: \begin{itemize}
\item A new component of moduli space in nine dimensions, which can be described equivalently as the asymmetric IIB orbifold with $C_0=1/2$ theta angle turned on, an $O8^+-O8^-$ compactification of IIA with a $C_1$ Wilson line, or F-theory on a certain non-orientable Bieberbach manifold;
\item A new component in eight dimensions, which is the dimensional reduction of the component above;
\item Two new components of moduli space in seven dimensions, described as type IIB on a free quotient of $T^3$ with an $SL(2,\mathbb{Z})$ bundle and non-trivial NSNS and RR 2-form profiles; we also provide novel, smooth descriptions of the known components of moduli space of low rank.
\item Potential discrete theta angles in the $SO(16)\times SO(16)$ non-supersymmetric, tachyon-free string.
\end{itemize}

To our knowledge, none of these components have been described before in the literature (the closest is the 4d model described in eq. (3.19) of \cite{Heidenreich:2016aqi}, which is a compactification of the new 9d theory we construct). They consist of new models of quantum gravity in high dimensions. As the known landscape of string compactifications in $d>7$ is not very large (prior to this paper, we knew of only four theories, modulo dualities, in nine dimensions, and three in eight), this expansion is significant. The theories we find are identical to their cousins without theta angle at the massless level, but differ significantly at the level of massive states.  Along the way, we clarify several minor puzzling features, such as why neither the rank 17 nor the rank 9 component in nine dimensions admit a theta angle, and the connection between bulk discrete theta angles and theta angles in the worldvolume theory of probe branes. We work out the exact duality orbit of the new theory, including all perturbative corners and how they fit in together, which we can do thanks to the large supersymmetry. We also clarify the M-theory construction of supersymmetric backgrounds preserving sixteen supercharges.

From the point of view of the Swampland Program \cite{Vafa:2005ui}, it is crucial to have a complete understanding of the Landscape, in order to minimize the risk of being misled by patterns that turn out to be only accidental. The  new nine-dimensional theory we find here is of relevance for the Swampland, since it is, to our knowledge, the first example of a quantum gravity with sixteen supercharges and an incomplete BPS string spectrum\footnote{This means that the model in \cite{Heidenreich:2016aqi}, which is a compactification of the 9d model with an incomplete string lattice, will also have an incomplete lattice of BPS strings. In that paper, the model was discussed as an example which violates BPS completeness for particles, but the string spectrum was not analyzed.}. It also satisfies only a $\mathbb{Z}_2$ sublattice version of the Weak Gravity Conjecture (WGC) for strings \cite{ArkaniHamed:2006dz,Montero:2016tif,Heidenreich:2016aqi}. This affects a number of proposed Swampland constraints, which were based on anomaly inflow on a BPS string of elementary charge. These papers should be read as obstructions to the existence of certain supergravities together with a complete BPS spectrum; but they may yet exist if they can be consistently be coupled to an incomplete BPS spectrum. At present we do not know how to figure out when this is possible and when it isn't; but the results we present here show that it is possible, at least for the case of the 9d $\mathcal{N}=1$ theories of rank 1.

Remarkably, we did not originally realize the existence of these theories by thinking hard about string constructions. Rather, all the supersymmetric theories presented in this paper were predicted by one of the authors in \cite{Hector-paper} using the framework of F/M-Theory with frozen singularities \cite{deBoer:2001wca,Tachikawa:2015wka,Bhardwaj:2018jgp} using lattice embedding techniques, and extrapolating the results to nine dimensions. This extended framework captures all known theories with 16 supercharges in dimensions 7,8 and 9, and predicts a few more. This paper presents the stringy embeddings of these new theories. It would be extremely interesting to provide a top-down derivation of the lattice embedding rules in \cite{Hector-paper} and figure out to what extent they can be generalized to setups with less supersymmetry. 

The paper is organized as follows: In Section \ref{sec:sethi} we introduce the Sethi string construction, and explain in detail why it does not work. In Section \ref{sec:new9d}, we use the same idea in nine dimensions to construct a new string theory. We study in detail its properties, spectrum, duality group, moduli space, as well as identify several related descriptions. We elaborate on the fact that the theory does not satisfy BPS completeness (although there is a complete lattice of non-BPS strings, in accordance with the Completeness Principle \cite{Polchinski:2003bq,Banks:2010zn}). In Section \ref{sec:comp} we describe the compactification of the previous theory to eight dimensions, and introduce the dual M-theory description in terms of a non-orientable Pin$^+$ Bieberbach manifold. In Section \ref{sec:new7d} we extend this classification to seven dimensions, discovering two new components of moduli space and a type IIB supergravity description of the known theories of low rank. In Section \ref{sec:nonsusy} we analyze potential discrete theta angles in nonsupersymmetric string theories. Finally, Section \ref{sec:conclus} presents our conclusions.

\section{The Sethi string and related discrete theta angles}\label{sec:sethi}
In \cite{Sethi:2013hra}, Sethi proposed a new string theory in 10 dimensions. The basic idea was quite simple: In the usual description of type I string theory as an $O9$ orientifold of type II, some of the IIB fields are projected out by the orientifold projection. In particular, under the action of $\Omega$ \cite{Dabholkar:1997zd,Tachikawa:2018njr},
\begin{equation} B_2\rightarrow -B_2,\quad C_4\rightarrow -C_4\quad C_2\rightarrow C_2,\quad C_0\rightarrow -C_0.\end{equation}
The fact that $B_2$ is projected out implies that type I strings can break. The unprojected $C_2$ is the 2-form that enters in the type I Green-Schwarz mechanism. We are interested in the RR axion. Since it is projected out, it is customary to take $C_0=0$. But \cite{Sethi:2013hra} pointed out that, since $C_0$ is periodic, $C_0\sim C_0+1$, the orientifold only forces
\begin{equation} C_0=-C_0+n,\quad n\in\mathbb{Z},\end{equation}
and as a result, both $C_0=0,1/2$ are allowed. Sethi then proposed that $C_0$ plays the role of a discrete theta angle, and $C_0=1/2$ gives a new string theory in ten dimensions, identical at the massless level but differing from type I at the level of massive states. However it was difficult to match this proposal across the duality web.  $C_0$ couples electrically to $D(-1)$ brane instantons. From a physical point of view, this means that, when turned on, this theta angle would provide a contribution to the path integral of 
\begin{equation} e^{\pi i\,n},\label{nssw}\end{equation}
where $n$ is the number of $D(-1)$-branes in the configuration. Hence the theta angle detects the number of $D(-1)$-branes modulo 2.

Interestingly, precisely this theta angle was considered by Witten in \cite{Witten:1998cd}. Type I branes are described in terms of KO-theory (associated to the $SO(32)$ gauge bundle of the $D9$ branes) and since $KO(S^{10})=\mathbb{Z}_2$, there is a ten-dimensional  instanton in type I string theory, obtained as a topologically non-trivial soliton in $\mathbb{R}^{10}$ protected by $\pi_9(SO(32))=\mathbb{Z}_2$. The pointlike limit of this instanton is precisely the type I ``-1''-brane described above. This can also be understood from the worldsheet tachyon condensation perspective.

In \cite{Witten:1998cd}, Witten discusses the possibility of a $\mathbb{Z}_2$-valued theta angle for which the aforementioned (-1)-brane picks up a minus sign. This is precisely the Sethi theta angle. The type I supergravity action (and indeed, the type I worldsheet perturbation theory as a whole) is invariant not just under $SO(32)$ gauge transformations, but also under their extension to $O(32)$. However, any $O(32)$ element in the connected component not containing the identity is anomalous. This reduces the actual, unbroken long-range gauge symmetry group of the theory to $SO(32)$ (or $\text{Spin}(32)/\mathbb{Z}_2$ when non-perturbative D-brane states are also included), matching its heterotic counterpart. For this reason, $O(32)$ is rarely discussed in the context of type I. However, the anomalous transformation can still be useful. The nature of the anomaly is such that, when the anomalous transformation is carried out, the path integral changes by an amount
\begin{equation} \exp(\pi i \mathcal{I}),\label{index0}\end{equation}
where $\mathcal{I}$ is a mod 2 index that counts the number of instantons described above, modulo 2. This is precisely the same factor turned on by the discrete theta angle in the path integral. Therefore, it was concluded in \cite{Witten:1998cd} that the later is actually unphysical. As a consequence, the Sethi string is completely equivalent to ordinary type I string.

To belabor the point, consider any supergravity correlation function to which  $\mathbb{Z}_2$ instantons contribute, in ordinary type I string theory. As explained in \cite{Witten:1998cd}, these instantons have an odd number $n$ of fermion zero modes. A non-vanishing correlation function to which these instantons contribute is then of the form
\begin{equation}\langle \lambda^n\rangle_{\text{Type I}}\sim\int D\lambda\, \lambda^n\, e^{-S},\label{amp0}\end{equation}
in $\mathbb{R}^{10}$, where $\lambda$ are the 10-dimensional gluinos. To compute this amplitude in the supergravity approximation, one would sum over all finite-action $\text{Spin}(32)/\mathbb{Z}_2$ bundles in $\mathbb{R}^{10}$, which include the instanton and its $O(32)$ transformed solution; although the $O(32)$ transformations are not a symmetry of the full theory, they can still be used to generate backgrounds from other backgrounds. Because of the anomaly, the measure $D\lambda$ in the partition function above picks a $-$ sign in the instanton background; as the gluino $\lambda$ also picks a $(-1)$ under the $O(32)$ transformation, the amplitude is invariant. 

Now consider the same amplitude in the Sethi string. It is identical to \eq{amp0}, except for an additional insertion of a factor \eq{index0},
\begin{equation}\langle \lambda^n\rangle_{\text{Sethi}}\sim\int D\lambda\, \lambda^n\, e^{-S} e^{\pi i \mathcal{I}}.\end{equation}
We see that the contribution of instantons flips a sign. But since $n$ is odd, this can be undone by simply redefining the gaugini by $\lambda\rightarrow-\lambda$ -- precisely the effect of the $O(32)$ transformation we described above.

These arguments are completely analogous to the usual story in QCD with massless quarks. Here there is also a (continuous) theta angle, but it drops out of any physical observables as it can be washed away by a chiral rotation of the quarks. Specifically, in a non-abelian gauge theory with a massless Dirac fermion, a chiral rotation
\begin{equation}\psi\,\rightarrow e^{i\varphi\gamma_5}\,\psi\end{equation}
induces via the chiral anomaly a shift in the action
\begin{equation} S\rightarrow S+ \int \varphi \, \text{tr}(F^2)\,,\end{equation}
and so if the original theory had a theta term $\int\theta  \text{tr}(F^2)$, a mere change of coordinates in the path integral would change it to $\varphi+\theta$. 

The argument above is only exactly true if the chiral symmetry is an exact symmetry of the action. For example, if there are higher-derivative terms such as  $(\bar{\psi}\psi)^n$ for $n>1$, the theory will still have massless quarks, but it will be impossible to perform the above manipulation. Analogously, what we established above is that the low-energy type I \emph{supergravity} does not have a $\mathbb{Z}_2$ discrete theta angle. For type I \textit{string theory}, however, the $O(32)$ symmetry is manifest at every order in string perturbation theory, and so the conclusion persists at all orders in string perturbation theory. If the $O(32)$ symmetry is really only broken by $(-1)$-brane instantons, the argument in \cite{Witten:1998cd} rules out the improved type I theta angle, at any order in string perturbation theory.

In a sense, the arguments above (which are just belaboring the point in \cite{Witten:1998cd}) do not settle the question completely. Although we have seen that the Sethi theta angle is trivial in supergravity, at every order in perturbation theory, and essentially in any observable we could compute, the argument does not exclude the possibility that there are exotic non-perturbative effects, not captured by any of the known effects in the type I string, that make the theta angle physical again. Here we just point out that if that was the case, the resulting discrete theta angle would in practice have none of the properties expected from $C_0$ (it would be invisible to D-branes, etc). It would be so different that perhaps we should think of it as a completely new discrete theta angle in type I altogether, unrelated to any of the type I supergravity fields. Indeed, the existence of such completely arbitrary discrete parameters is very difficult to rule out; for a IIB example, consider the alternative anomaly cancellation mechanisms described in \cite{Debray:2021vob}. 

Another way to interpret the above is that type I string theory can somehow be viewed as an $O(32)$ theory (or more precisely, a $\text{Pin}$ lift of it) spontaneously broken to $\text{Spin}(32)/\mathbb{Z}_2$ by the vev of $C_0$, which is charged in the determinant representation\footnote{This is the 1-dimensional representation of $O(n)$ that sends elements in the connected component to $+1$ and those in the disconnected component to $-1$.}.  Another point is that an $O(32)$ reflection must flip the chirality of the heterotic spinor; it was observed in footnote 8 of \cite{Bergman:2000tm} that the non-BPS, unstable $D8$ brane of type I is precisely the domain wall between these two vacua. It would be interesting to see if this perspective can be pursued any further.

If $C_0$ is really unphysical, we should be able to see this in the worldvolume of branes as well. Consider first the worldvolume theory of a type I $D5$ brane, which can also be viewed as a small type I instanton \cite{Witten:1995gx,Witten:1998cd}. This brane has $SU(2)$ gauge fields at low energies, and the worldvolume theory admits a discrete theta angle associated to $\pi_5(SU(2))=\mathbb{Z}_2$. This charge allows one to construct a $D(-1)$ in the worldvolume of a $D5$ brane. However, the theory also has an $SO(32)$  flavor symmetry \cite{Witten:1998cd}, corresponding to the bulk $SO(32)$ group, which remains unbroken in the small instanton limit; There are hypermultiplets transforming in the $(\mathbf{32},2)$ of $SO(32)\times SU(2)$. Just as above, the perturbative flavor symmetry is really $O(32)$; the worldvolume fermions are anomalous under the $O(32)$ transformation in such a way so as to render the theta angle unobservable, essentially for the same reasons as in the bulk.

The would-be type I discrete theta angle can also be studied in different corners of the duality web. We will now see that our proposal that it is unphysical fits what we know from other sources. For instance, \cite{Bergman:2013ala} discusses a theta angle in type I' theory -- the T-dual to a circle compactification of type I (see e.g. \cite{Polchinski:1995df,Johnson:2000ch}, described by an interval compactification with $O8^-$ planes at the endpoints whose RR tadpole is canceled by $16$ mobile $D8$ branes. Much like the Sethi theta angle involves turning on $C_0$ as allowed by orientifolds, the discussion in \cite{Bergman:2013ala} is based in the observation that the $O8^-$ compactification involved in type I' projects out $\int_{S^1} C_1$, the holonomy of the RR field $C_1$ on the covering circle, to either zero or $1/2$. This would constitute a theta angle in nine dimensions; the authors construct explicitly the domain wall between the $\int_{S^1} C_1=0$ and $\int_{S^1} C_1=1/2$ phases, which is an unstable D7 brane \cite{Bergman:2013ala}. 

However, since this angle is T-dual to the Sethi theta angle, in the presence of D8 branes it too must be unobservable. Indeed, \cite{Bergman:2013ala} describes how the addition of $D8$ branes provides additional massless modes that make the theta angle unobservable, in the same way as above.  The $C_0$ theta angle corresponds to the holonomy  $\int_{S^1} C_1=0$; the $O(32)$ gauge transformation that exhibits the anomaly acts on the positions of the $D8$ branes as a sign flip. 

This highlights an interesting subtlety in type I'. The customary description of this background is as an interval $S^1/\mathbb{Z}_2$, where one must add $D8$ branes \cite{Johnson:2000ch}. Consider the covering $S^1$, described as the real line (parametrized by a coordinate $x$) subject to the identification $x\sim x+1$. The $\mathbb{Z}_2$ acts by sending $x\rightarrow -x$ so there are orientifold planes at $x=0,1/2$. The 16 $D8$ branes are located at particular points in this interval; in the double cover, a $D8$ brane at $x=a$ is accompanied by its orientifold image at $x=-a$.

From this geometric description, it may seem that putting a $D8$ brane at $x=a$ or at $x=-a$ is immaterial, as the corresponding orientifold images will be located at $x=-a$ or $x=a$ respectively. The actual situation is however a bit subtler. Consider the point of type I' moduli space dual to type I on the circle with no Wilson lines, where all $D8$ branes sit at the origin generating an $SO(32)$ gauge algebra. The scalars in the vector multiplets live in the adjoint of $SO(32)$, and so their vev space can be parametrized by Cartan generators.  We can describe a general element in the Cartan via the matrix
\begin{equation} v=\left(\begin{array}{cccccc}
0&a_1&\multicolumn{2}{c}{\multirow{2}{*}{$\quad\cdots$}}&0&0\\
-a_1&0&&&0&0\\
\multicolumn{2}{c}{\multirow{2}{*}{$\vdots$}}&\multicolumn{2}{c}{\multirow{2}{*}{$\ddots$}}&\multicolumn{2}{c}{\multirow{2}{*}{$\vdots$}}\\
&&&&&\\
0&0&\multicolumn{2}{c}{\multirow{2}{*}{$\quad\cdots$}}&0&a_{16}\\
0&0&&&-a_{16}&0
\end{array}\right)\end{equation}
where the $a_i$ parametrizes the position of the $i$th $D8$-brane. Focusing on a single $2x2$ block, we see that $a_1$ and $-a_1$ do not describe the same configuration, even after conjugation by the $SO(32)$ transformations. However, they are related by an $O(32)$ transformation:
\begin{equation}\left(\begin{array}{cc}-1&0\\0&1\end{array}\right)\cdot \begin{pmatrix}\,0&\,\,a_1\\-a_1&\,0\end{pmatrix} \cdot\left(\begin{array}{cc}-1&0\\0&1\end{array}\right)^{-1}=\begin{pmatrix}\,0&\,\,-a_1\\a_1&\,0\end{pmatrix}.\end{equation}
This is the same transformation that was shown to be anomalous in ten dimensions. In ten dimensions, the relevant instanton was associated to $\pi_9(SO(32))=\mathbb{Z}_2$; the dimensional reduction of this instanton is now related to  $\pi_8(SO(32))=\mathbb{Z}_2$ in the worldvolume theory of $D8$ branes on top of the orientifold. Just as above, this transformation is anomalous.

\begin{figure}[hbtp!]
\centering 
    \includegraphics[width=0.6\textwidth]{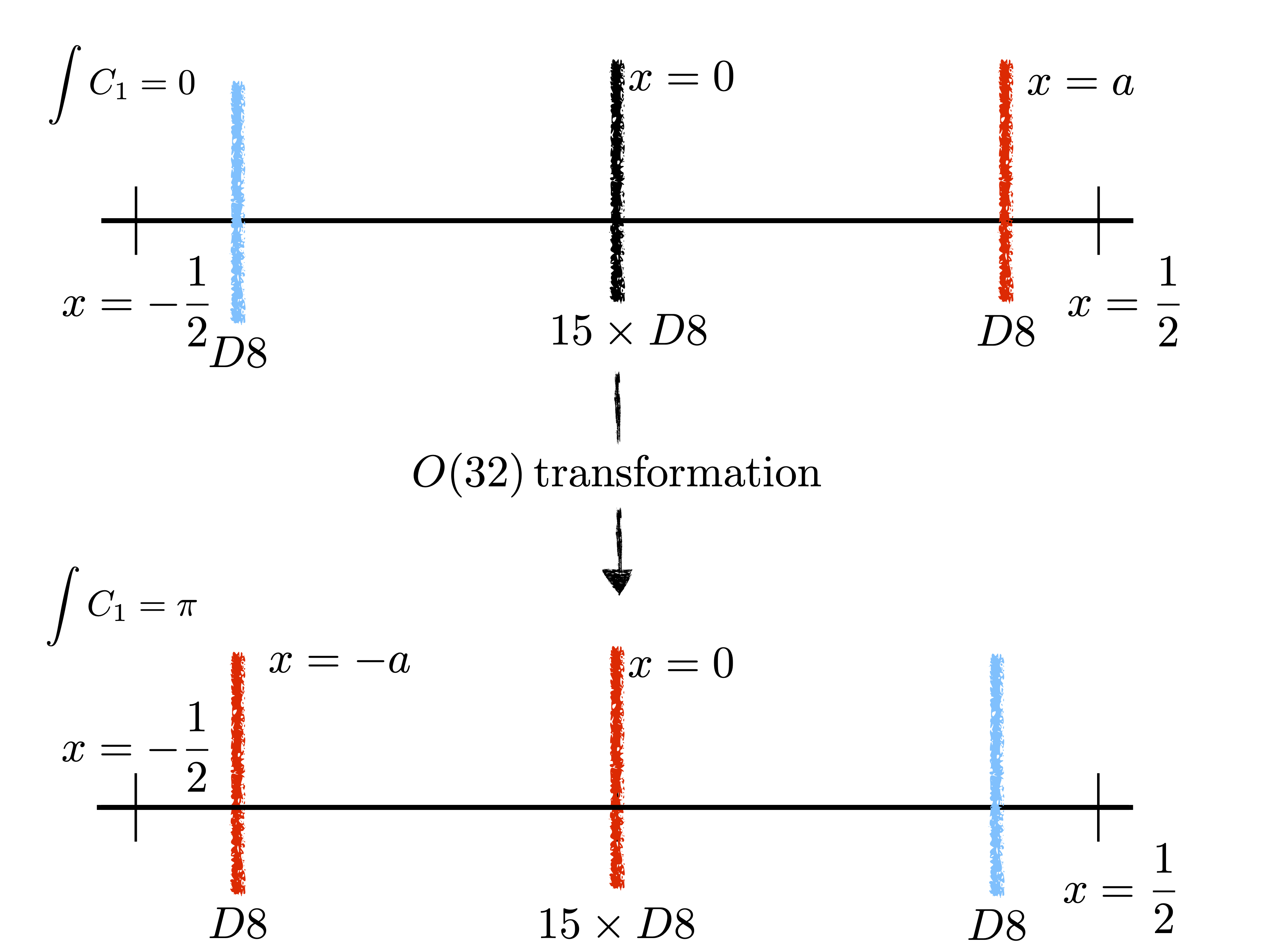}
    \caption{The type I' theory describes $D8$ branes on an interval, depicted here as the real line $\mathbb{R}$ subject to periodic identifications. $O8^-$ planes are found at $x\in\mathbb{Z}/2$. The top section of the figure describes 15 $D8$ branes sitting at the origin and a single $D8$ brane moved to $x=a$. There is no Wilson line. Thanks to the anomaly described in the main text, acting with an $O(32)$ transformation flips the position of the $D8$ brane to $x=-a$ \emph{and} introduces a discrete Wilson line. Since the brane at $x=-a$ can also be obtained simply by moving the $D8$ to the left, we discover that the discrete Wilson line is not an invariant notion.  }
    \label{f0}
\end{figure}

 We can now make the precise statement: Type I' is invariant under the operation of flipping the position of an odd number of $D8$ branes, together with the introduction of a discrete Wilson line $\int_{S^1} C_1=1/2$ at the same time (see Figure \ref{f0}). Thus, there is no discrete theta angle for type I'; it can be made to appear or disappear simply by a choice of coordinates in moduli space.
 
The existence of this anomaly has interesting consequences. Consider type I', at the point in moduli space described above,which is T-dual to type I on a circle without Wilson line. One has 32 $D8$ branes sitting on top of an $O8^-$ at this point, and there are gauge bosons in an $\mathfrak{so}(32)$ Lie algebra. Let us choose $[0,1/2]$ as the fundamental domain, and have the $32$ D8 branes sitting at $x=1/2$. Consider moving a single $D8$ brane from $x=1/2$ to $x=-1/2$. Since at this point there are also 32 coinciding D8 branes, we will have a configuration with $\mathfrak{so}(32)$ Lie algebra, and the naive expectation would be that this is in fact the same configuration we started with, corresponding to type I on a circle. However, this is not correct. As mentioned above, this configuration is equivalent to having $32$ D8 branes at $x=1/2$, together with a Wilson line $\int_{S^1} C_1=1/2$ turned on. This is not the same point in moduli space we started with, which had $\int_{S^1} C_1=0$. 

Thus, moving a $D8$ brane across the type I' interval does not result in the type I point, but rather it takes us to a different point in moduli space with a $\mathfrak{so}(32)$ algebra. Indeed, there are two separate loci in the moduli space of type I' (the unique moduli space of rank 17 nine-dimensional quantum gravity) with this symmetry algebra \cite{Narain:1986am,Font:2020rsk}. In the heterotic perspective, they correspond to different choices of Wilson line preserving the $\mathfrak{so}(32)$ symmetry. The two heterotic vacua also differ in their spectrum of massive states. The ordinary heterotic vacuum has a spectrum of perturbative excitations consistent with the global form of the gauge group $\text{Spin}(32)/\mathbb{Z}_2$; by contrast, the other point contains vectors and spinors of opposite chirality, which are also charged under the graviphoton. In both cases, taking into account the vector in the gravity multiplet, the actual gauge group is 
\begin{equation}\text{Spin}(32)\times U(1),\end{equation}
and a particular value of the radius it enhances to a non-abelian group. In the ordinary heterotic vacuum, the group $\text{Spin}(32)/\mathbb{Z}_2 \times U(1)$ enhances to $\text{Spin}(32)/\mathbb{Z}_2 \times SU(2)$; while in the other locus, the enhancement is to Spin$(34)$.

Since the gauge algebra is invariant when we change the coupling (the scalar in the gravity multiplet), a similar picture must be true in the type I picture we began with. Let us first see how to recover the $SU(2)$ enhancement in the ordinary heterotic vacuum, which is dual to the at the ordinary type I point. This is just the T-dual of type I on a circle with no Wilson line, so the algebra is $\mathfrak{so}(32)$. We now move to small radius to find the self-dual point. There is an upper bound on the value of the dilaton, $g\geq 1/R$, which comes from demanding that the gauge coupling at $x=0$ does not become negative \cite{Polchinski:1995df}. When reaching this limit, we expect that $D0$ branes become massless at $x=0$ \cite{Morrison:1996xf,Bergman:1997py,Bachas:1997kn,Aharony:2007du}; this follows from an analysis of the $D0$ brane quantum mechanics on the interval \cite{Bachas:1997kn}, taking into account the warped metric and non-trivial dilaton profile. At a certain point, these $D0$ branes become massless, enhancing the $U(1)$ factor to $SU(2)$, but producing no enhancement of the $\text{Spin}(32)/\mathbb{Z}_2$ factor.  

What we have just described is the type I' explanation of why there is no enhanced symmetry when type I string theory is compactified on a circle of stringy size. Let us see however how this changes when the discrete Wilson line $\int_{S^1} C_1=1/2$ is turned on. As explained above, from the heterotic analysis we expect an enhanced $\text{Spin}(34)$ symmetry; how can we see that this is the case from the type I' perspective? Turning on a Wilson line $\int_{S^1} C_1=1/2$  means that $D0$ branes have antiperiodic, rather than periodic boundary conditions on the covering circle. Their spectrum is different, and it is natural to expect that the localized massless $D0$ branes that existed before are now not present. Instead, at the locus where the coupling vanishes, the $O8^-$ at strong coupling emits an additional $D8$ brane \cite{Morrison:1996xf}; one can formally push to regimes below $g>1/R$, where the coupling constant becomes the position of this effective $D8$ brane on the interval. The furthest it can go is to $x=1/2$, where it enhances the symmetry to $\text{Spin}(34)$.

Prior literature explained that $O8^-$ planes can either have massless $D0$ branes on top of them, or non-perturbatively emit $D8$ branes in some cases \cite{Morrison:1996xf,Bachas:1997kn,Aharony:2007du,Cachazo:2000ey} (see \cite{Bachas:1997kn} for an alternative description that does not involve $D8$ branes; we will stick with the $D8$ brane description, but the description of the Spin(34) locus of moduli space in that reference is consistent with ours). It was not clear under which conditions can each of these phenomena occur. The above analysis shows that the distinguishing factor is the discrete theta angle, and  leads to a simple picture for what happens:\begin{itemize}
\item An $O8^-$ with $\int_{S^1} C_1=0$ has massless $D0$ branes stuck there; these can enhance the symmetry to exceptional groups if additional $D8$ branes are present.
\item An $O8^-$ with $\int_{S^1} C_1=1/2$ cannot have $D0$ branes becoming light; on the other hand, it can non-perturbatively emit an additional $D8$ brane, which provides grounds for further enhancement of the symmetry group.
\end{itemize}

This distinction can also be seen at the level of probe instantonic 4-branes \cite{Morrison:1996xf,Hamada:2021bbz,Bedroya:2021fbu}. A probe $D4$ brane located at a point in the interval can be used to probe the structure of the different singularities in the compactification. The distinction above is mirrored in the worldvolume theory of a brane probing a non-perturbative $O8^-$ plane: 
\begin{itemize}
\item A $D4$ brane probing an $O8^-$ with $\int_{S^1} C_1=0$ yields an $E_1$ SCFT, with $U(1)$ global symmetry \cite{Morrison:1996xf,Hamada:2021bbz,Bedroya:2021fbu};
\item On the other hand, an $O8^-$ with $\int_{S^1} C_1=1/2$ realizes an $\tilde{E}_1$ SCFT, with no global symmetry.
\end{itemize}
The relationship between these SCFT's described in \cite{Morrison:1996xf} precisely mimicks the discussion above. In particular, both the $E_1$ and $\tilde{E}_1$ SCFT's arise from RG flow of the $E_2$ SCFT, depending on the sign of a certain mass term. We can now explain this in the brane picture: the $E_2$ SCFT  is the worldvolume theory of a $D4$ probing a $O8^-+D8$ system. The sign of the mass term corresponds to the position of the $D8$ brane. The fact that different signs of the $D8$ brane position lead to different IR physics is precisely the feature described above, and related to the anomaly of the gauge transformation in the disconnected part of the gauge group. As is well-known, the $E_1$ SCFT leads to a $SU(2)$ gauge theory at low energies on a generic point of its Coulomb branch; while the $\tilde{E}_1$ SCFT is described by $SU(2)$ gauge theory with a discrete theta angle turned on. It is natural to identify the spacetime angle $\int_{S^1} C_1=1/2$ with the worldvolume discrete theta angle, as was proposed in \cite{Bergman:2013ala}.

Finally, we also comment on discrete theta angles in other theories. There is  no discrete theta angle either in the rank 9 (CHL) component of moduli space, but the reason is different. This component of moduli space can be described as an $O8^-+O8^0$ orientifold with $8$ $D8$ branes \cite{Keurentjes:2001cp,Aharony:2007du}. The $O8^0$ is a slightly exotic orientifold plane, introduced by \cite{Keurentjes:2001cp,Aharony:2007du} to explain agreement with the M-theory picture of the rank 9 and one of the rank 1 components of moduli space. It corresponds to compactifying M-theory on a cross-cap ($\mathbb{RP}^2$ minus a point) geometry. We claim that in any compactification involving $O8^0$ planes, the holonomy $\int C_1$ is frozen to a non-zero value (in fact, this claim already appears in \cite{Keurentjes:2001cp}). This means that a $D0$ brane bouncing back off an     $O8^0$ always picks up a factor of $-1$. This can be seen explicitly by looking at the local geometry of an $O8^0$ plane realized e.g. as the dimensional reduction of a Mobius strip \cite{Keurentjes:2001cp,Aharony:2007du}; the definition involves orientation reversal, and so it flips KK momentum (or equivalently, $D0$ brane charge). Thus, there is no discrete theta angle in the CHL string either, or in the rank 1 component of moduli space obtained as M-theory on the Klein bottle.

\section{New string theories in nine dimensions from theta angles}\label{sec:new9d}
Although the  construction in \cite{Sethi:2013hra} does not quite work in 10 dimensions, the reason for the failure is a technicality. In this Section we will see how the idea can be actually quite prolific and lead to the discovery of new string compactifications with sixteen supercharges. We will start with the example with the closest resemblance to Sethi's construction; and later will discuss other examples and the ramifications across the duality web.

\subsection{The rank 1 Sethi string}\label{subsec:10dsethi}
Type IIB string theory has two perturbative symmetries, dubbed $(-1)^{F_L}$ and $\Omega$ \cite{Dabholkar:1997zd,Tachikawa:2018njr}. Consider a compactification of IIB on an $S^1$ with a discrete Wilson line for $\Omega$ along the circle. This is a 9d $\mathcal{N}=1$ theory, known as the Dabholkar-Park (DP) background \cite{Dabholkar:1996pc,Aharony:2007du}. It describes one corner of one of the two known components of the moduli space of 9d $\mathcal{N}=1$ theories.

The $\Omega$ symmetry flips the sign of $C_0$, so just like in type I,  the RR axion is projected out in the DP background. However, we can now play the same game as Sethi did in ten dimensions: It is consistent to set $C_0=0$ or 
\begin{equation} C_0=1/2\quad \text{in the DP background}.\label{dp99}\end{equation}
We claim that, unlike in 10 dimensions, the discrete theta angle is physical, cannot be gauged away, and that \eq{dp99} describes a new string theory in nine dimensions with sixteen supercharges. We will provide more evidence below, but perhaps the simplest,  argument is the fact that, unlike in type I, $D(-1)$ instantons are not projected out\footnote{Although it is worth noting that the holonomy means that moving a $D(-1)$ a full turn around the circle turns it into a $\overline{D(-1)}$, just like when moving around an Alice string \cite{Schwarz:1982ec,Preskill:1990bm,Fukuda:2020imw}.}, so they are sensitive to the precise value of $C_0$.

\begin{figure}[hbtp!]
\centering 
    \includegraphics[width=0.6\textwidth]{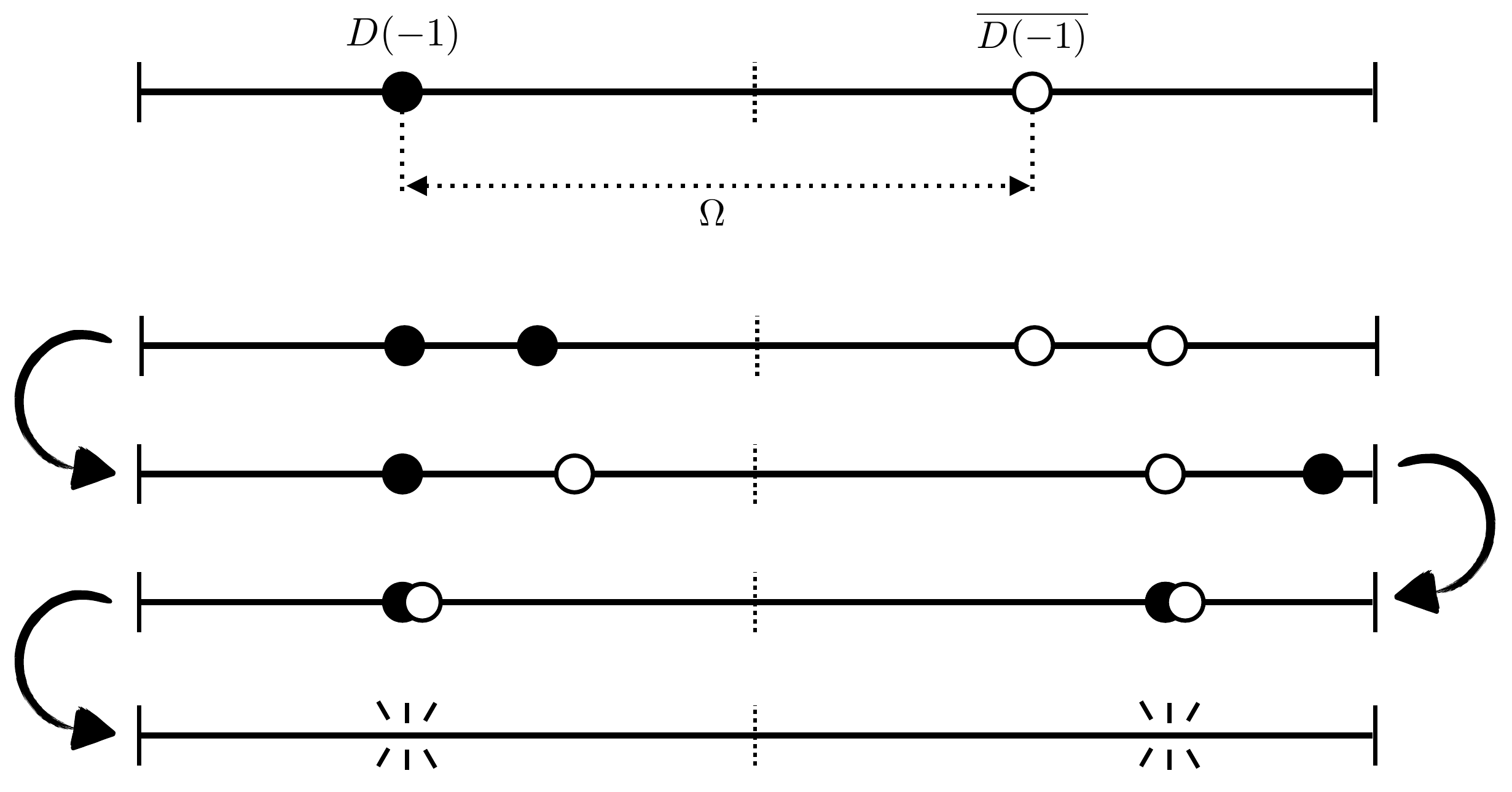}
    \caption{Picture of the DP background via its covering $S^1$. The two ends of the interval must be identified, forming a circle. The  DP background is obtained by restricting to configurations invariant under a half-shift combined with an action of $\Omega$. In the top panel of the Figure, we illustrate one such configuration, consisting of a $D(-1)$ brane (black dot) and an $\overline{D(-1)}$ (white-filled dot) in a shifted position. This configuration couples to the $\mathbb{Z}_2$ field $C_0$, and is stable. Putting two such objects together allows for their decay by positioning the antibrane of one of the pairs on top of the brane of the other, as illustrated in the successive steps in the bottom panel.}
    \label{f1D7}
\end{figure}

One crucial difference is that, in accordance with the fact that $C_0$ is a $\mathbb{Z}_2$-valued theta angle, the $D(-1)$ instantons are $\mathbb{Z}_2$-charged as well in the DP background, meaning that two of them can be smoothly deformed to the vacuum.   One way to see this in physical terms is to consider that if one has e.g. a $D(-1)$  sitting at a particular point on $S^1$, its image in the double cover includes a $\overline{D(-1)}$ in the antipodal point. Put now two of these $D(-1)$ branes and it is possible to move one image into the anti-image of the other, annihilating the whole configuration, as depicted in Figure \ref{f1D7}.

The theory we describe here does not appear anywhere in the two known connected components of the moduli space of 9d $\mathcal{N}=1$ theories of rank 1 \cite{Aharony:2007du}; Sethi's construction has succeeded in producing a genuinely new string theory in nine dimensions. 

Other $D$-branes are also sensitive to the value of this discrete theta angle, since they couple to it via the CS worldvolume coupling \cite{Polchinski:1995mt,Douglas:1995bn,Johnson:2000ch}
\begin{equation} \int_{\text{Dp-brane}} C_0 \left[\sqrt{\hat{A}(R)} \text{ch}[F]\right]_p.\label{ew200}\end{equation}
The branes see this as a discrete theta angle in their worldvolume. $D3$ and $D7$ branes wrapped on the $S^1$ are projected out by the Wilson line in the DP background; However, $D1$ and $D5$ branes both wrapped and unwrapped along the $S^1$ survive as states in the 9d theory. This corresponds to the fact that $C_2$ is unaffected by the $\Omega$ action, and so it yields both a 2-form in 9d (the 2-form in the 9d $\mathcal{N}=1$ gravity multiplet \cite{Aharony:2007du,Kim:2019ths}) and a 1-form (one of the two vectors in a 9d $\mathcal{N}=1$ theory of rank 1, the other being the KK photon). The unwrapped $D1$ brane and wrapped $D5$ branes correspond to the electrically and magnetically charged objects with respect to the 2-form; the wrapped $D1$ and unwrapped $D5$ correspond to the electrically and magnetically charged objects under the vector. For the $D1$ and $D5$ branes the coupling \eq{ew200} takes the form 
\begin{equation}   \int_{D1} C_0 F,\quad \int_{D5}\frac{C_0}{6} F \wedge\left(F^2+\frac{1}{16}\text{Tr}(R^2)\right).\label{e23}\end{equation}
These couplings manifest themselves into physical properties of the states that can be constructed as solitons of the worldvolume theories of the branes. Most outstandingly, the tension of the $D1$ brane is sensitive to the value of the RR axion \cite{Witten:1995im,Schwarz:1995du,Schwarz:1995dk}. This is because the RR theta angle shifts the quantization condition for string winding number, in such a way that the ground state of a $D1$ acquires a bit of fundamental string charge, similarly to the Witten effect in 4d gauge theory \cite{Witten:1979ey}. The correct tension formula for the tension of a $(p,q)$ string in 10d Planck units \cite{polchinski1998string} means that the tension of a single $D1$ is
\begin{equation} T_{D1}=\frac{1}{2\pi\alpha'}\sqrt{\frac14+\frac{1}{g_s^2}},\label{bpst5}\end{equation}
which contrasts with the tension of the $D1$ brane at $C_0=0$. This clearly establishes that the theta angle we are discussing is physically meaningful, unlike the examples in the previous Section. To find the spectrum of actual states one will have to consider the full state comprised of the $D1$ brane and its images under the DP action, as we will do below.

 As will be described in more detail in Subsection \ref{subsec:dual}, turning on the discrete theta angle is more properly described as a circle compactification of IIB with a holonomy of $\Omega\, T$, where $T$ is the usual generator of the modular group that shifts the RR axion by 1. Including $T$ is necessary (and equivalent) to turning on the theta angle, and has important effects on the spectrum of charged objects. The action on the RR and NS-NS 2-form fields is\footnote{Note that the action described in the main text is not an element of $SL(2,\mathbb{Z})$, but rather it lies in $GL(2,\mathbb{Z})$. The actual duality group of IIB is closer to the latter than the former (see e.g. \cite{Tachikawa:2018njr}); the perturbative IIB symmetries $\Omega$ and $(-1)^{F_L}$ that we use here are precisely responsible for the extension from $SL(2,\mathbb{Z})$  to $GL(2,\mathbb{Z})$.}
\begin{equation}\left(\begin{array}{c}C_2\\B_2\end{array}\right)\,\rightarrow\, \left(\begin{array}{cc}1&1\\0&-1\end{array}\right)\left(\begin{array}{c}C_2\\B_2\end{array}\right)=\left(\begin{array}{c}C_2+B_2\\-B_2\end{array}\right).\end{equation} 
The field combination $B_2+2C_2$ is invariant, and thus yields the 2-form of the nine-dimensional gravity multiplet, while $B_2$ picks up a $-$ sign. The coupling of a $(p,q)$ string to the 2-forms can be written as
\begin{equation}\int p B_2+qC_2=\int \frac{q}{2} (B_2+2C_2)+\left(p-\frac{q}{2}\right)B_2\label{e32co}.\end{equation}
In this basis, the action of the holonomy $\Omega T$ leaves the first charge invariant, while flipping the sign of the second. A $D1$ brane ($q=1$) is charged under the nine-dimensional 2-form, but also under the torsional field $B_2$. As a result, it is not a BPS object. This can also be seen in the covering space description: A $D1$ brane is not invariant under $\Omega T$, and the image $(1,-1)$-brane is not mutually BPS with the $D1$ brane. On the other hand, a $(p,q)=(1,2)$ string is invariant under $\Omega T$, and is actually a BPS object. 

The physical, long-range string charge of the $D1$ brane is $1/2$, while that of the $(1,2)$ string is 1. This means that, unlike every other know example of quantum gravity with 16 supercharges,  BPS strings only exist for a sublattice of index 2 of the allowed set of charges for the two-form. This new string theory therefore provides a counterexample with sixteen supercharges to the BPS completeness conjecture of \cite{Kim:2019vuc}, which posited that, with enough supersymmetry, there are BPS strings with every possible value of the charge\footnote{Of course, there are still non-BPS strings for every value of the charge: the usual Completeness Principle \cite{Polchinski:2003bq,Banks:2010zn} is satisfied.}. This has significant implications for the Swampland program; for instance, the arguments in papers like \cite{Kim:2019vuc,Lee:2019skh,Lanza:2019xxg,Lanza:2020qmt,Kim:2019ths,Katz:2020ewz,Angelantonj:2020pyr,Cvetic:2020kuw,Tarazi:2021duw,Cvetic:2021vsw} may have to be revisited.

All of the above can be verified directly from the string tension formula in Planck units \cite{polchinski1998string}
 \begin{equation} T_{(p,q)}=\frac{\vert p+q\tau\vert}{\sqrt{\text{Im}(\tau)}}=\sqrt{\left(p^2-pq+\frac{q^2}{4}\right) g_s+\frac{q^2}{g_s}},\label{tpq}\end{equation}
 where in the second step we have substituted $\tau=-\frac12+ i/g_s$.  For even RR charge $q$, the tension is minimized by the $(-q/2,q)$ string. For odd $q$, the tension is instead minimized by the string with charges 
 \begin{equation}(p,q)=\left(\frac{q\pm1}{2},q\right)\end{equation}
 These two strings are classically degenerate; we expect this degeneracy is lifted by quantum corrections, and that there is a single linear combination of lowest tension. This string is not BPS, but it is stable, since it is the lightest string in its charge sector.  The charge-to-tension ratio of the strings with odd charge, normalized to the charge-to-tension ratio of the BPS string, is
 \begin{equation}\frac{\xi_{\text{Half-integer q}}}{\xi_{\text{BPS}}}=\frac{1}{\sqrt{1+\frac{g_s^2}{4q^2}}},\end{equation}
 so that odd-charge strings are indeed subextremal.

These considerations also have ramifications for the spectrum of electrically charged particle BPS states, which are obtained as winding states of the BPS $(1,2)$ string wrapping the $S^1$ with KK momentum. Just as for the strings, the BPS particles only populate a sublattice of index two in the whole charge lattice, providing an example of a sublattice WGC in high dimension \cite{Montero:2016tif,Heidenreich:2016aqi}. 

One could also view these results as suggesting that the Sethi string is different to type I, contrarily to what was argued in Section \ref{sec:sethi}\footnote{We thank Ben Heidenreich for raising this point.}. Viewing the Sethi string as an orientifold of IIB by $\Omega T$, one could consider the state obtained by acting with the orientifold projection on a IIB $D1$ string and its $(1,1)$ image. If this had fractional charge, then the Sethi string would be physically different from type I; but it turns out to have the same charge as an ordinary BPS brane. The reason for this discrepancy is that the ordinary type I string is the orientifold image of a \emph{single} IIB D1 string \cite{Gimon:1996rq}\footnote{This is in contrast to the situation for $D5$ branes, where two are required since the orientifold action is symplectic \cite{Gimon:1996rq}. Relatedly, we do find a full charge }. This is required for consistency with Dirac quantization \cite{Gimon:1996rq}. As a result, with theta angle turned on, we expect the $(2,1)$ IIB string to project to the fundamental charge 1-brane after taking the quotient. This has the same physical charges as a $(0,1)+(1,1)$ stack. By contrast, in the AOB background with theta angle, a single $(2,1)$ would not be invariant; either two $(2,1)$'s or a $(0,1)+(1,1)$ are, leading to the sublattice described above.

The discrete theta angle also has effects on the $D5$-brane, where the non-zero expectation value for $C_0$ turns on a theta angle for the $U(1)$ worldvolume theory. Unlike for $D1$'s, the tension of the D5-brane (which is BPS) is non-renormalized\footnote{We would expect this effect for the NS5 5-brane, but this is projected out by the $\Omega$ Wilson line. In terms of kinematics, D5 branes behave as fundamental strings, and NS5 branes act as D1-strings \cite{Witten:1995im}.}, but there are worldvolume effects, such as changing the fermion parity of instanton strings in the worldvolume of the brane; reference \cite{Bergman:2013ala} studied this in a dual description in terms of $D4$ branes.

Finally, the theory also contains $D3$ and $D7$ branes, corresponding to ordinary $D3$ and $D7$ branes unwrapped on the circle (they cannot be wrapped on the circle due to the $\Omega$ action). Just like $C_0$, the corresponding RR fields $C_4$ and $C_8$ are projected out down to a $\mathbb{Z}_2$ subsector; the DP background therefore has discrete $\mathbb{Z}_2$ 3-form and 8-form fields, respectively. Such discrete fields have been noticed recently in a different 9d $\mathcal{N}=1$ theory, and take part into a beautiful mechanism to cancel anomalies of M-theory on a Klein Bottle \cite{Lee:2022spd}. Because the $D3$ and $D7$ are charged under $\mathbb{Z}_2$-valued fields, they are themselves $\mathbb{Z}_2$-charged.  The argument is similar to the one given for $D(-1)$-branes at the beginning of this Section.

The $\mathbb{Z}_2$ $D7$-brane is actually the domain wall interpolating between the values $C_0=0,\pi$ of the discrete theta angle, predicted by the cobordism conjecture \cite{Bergman:2001rp,McNamara:2019rup}. To see this, we will compute the axio-dilaton profile sourced by a the $D7$ and $\overline{D7}$ branes located at antipodal points in the covering circle. A $D7$ brane located at $z=z_0$ sources an axio-dilaton profile \cite{Weigand:2018rez}
\begin{equation} d\tau_{D7}(z)=\frac{1}{2\pi i\,(z-z_0)}\end{equation}
away from its sources, while the profile of a $\overline{D7}$ brane at the same location is minus the complex conjugate of the above. We will now describe $S^1\times \mathbb{R}$ via coordinates coordinate $z=x+i y$, and identification $y\sim y+1$. The profile we look for can be obtained by adding up the contributions of infinitely many $D7$ branes located at $\mathbb{Z}\, i$, and infinitely many $D7$ branes located at $\left(\mathbb{Z}+\frac12\right)i$, plus constants to regularize the sum that will not affect the field strengths. We obtain
\begin{equation}2\pi i\, d\tau=\frac{1}{z}+\sum_{n\neq0} \frac{1}{z-ni} +\frac{1}{ni}+\sum_{n} \frac{1}{z-\left(n+\frac12\right)i}+\frac{1}{\left(n+\frac12\right)i}=\pi\left(\coth[\pi z] -\tanh(\pi\bar{z})\right),\end{equation}
which integrates to
\begin{equation}\tau= \frac{1}{2\pi i} \log\left(\sinh(\pi z)\cosh (\pi \bar{z})\right).\end{equation}
The profile of $C_0$ is just the real part of $\tau$ above. One can check explicitly that $C_0\rightarrow 0$ at $x\rightarrow\infty$, while $C_0\rightarrow \pm\frac12$ when $x\rightarrow -\infty$, as advertised, showing that the non-BPS $D7$ brane is the domain wall we were looking for \cite{Bergman:2001rp}.

We wrap up this Subsection by pointing out that all of these discrete theta angles and fields are actually implicitly predicted by the K-theory description of branes \cite{Witten:1998cd} and its generalization including bundle involutions carried out in \cite{Bergman:1999ta,Bergman:2001rp}. In the case at hand, we must consider a stack of $n$ $D9-\overline{D9}$ pairs on the DP background; as described in \cite{Bergman:1999ta,Bergman:2001rp}, the appropriate K-theory classifying these backgrounds is the theory called $KR(X^9\times S^{2,0})$ in \cite{Atiyah:1966qpo}, where $X_9$ is the 9-dimensional spacetime. This K-theory is identical to another theory, called $KSC$ which is 4-periodic and has the groups \cite{10.2307/71980}
\begin{equation} \begin{array}{c|ccccccccc}m&1&2&3&4&5&6&7&8&9\\\hline KSC(S^m)&\mathbb{Z}_2&0&\mathbb{Z}&\mathbb{Z}&\mathbb{Z}_2&0&\mathbb{Z}&\mathbb{Z}&\mathbb{Z}_2\end{array}\end{equation}
which matches the branes we found before on more physical grounds. This was first noticed in \cite{Bergman:1999ta}, where the DP background is referred to as type $\tilde{I}$. On top of these, of course, there are also discrete fields and branes coming from the NS-NS sector; a 2-form field coming from the $B$-field, and its dual $\mathbb{Z}_2$ 5-form. These are not captured by a K-theory description. 

\subsection{The AOB background with theta angle}\label{subsec:aobth}
The technique used in the previous Subsection to exhibit a theta angle for the DP background also works for a background with the ``$S$-dual'' holonomy of $(-1)^{F_L}$. Such a background was constructed in \cite{Hellerman:2005ja}, and receives the name of the ``Asymmetric Orbifold of IIB'', or AOB for short. The action of $(-1)^{F_L}$ on $\tau$ is the same as that for $\Omega$, and so, here too we can turn on a theta angle $C_0=1/2$. 

The resulting theta angle is also genuine, since just as in the previous Subsection, it is detected by $\mathbb{Z}_2$ $D(-1)$-branes, although its physical effects look very different from those of the theta angle for the DP background. The main reason for this difference is that, when the theta angle is turned off, the $(-1)^{F_L}$ orbifold projects out the RR fields $C_2,C_6$ (in addition to $C_0,C_4$), and keeps the NS-NS fields $B_2, B_6$. Correspondingly, the only branes with $\mathbb{Z}$-valued charges are fundamental strings and NS5-branes, both wrapped and unwrapped on the $S^1$. Before, we could see the effects of the discrete $C_0$ angle on the worldvolume of $D1$ and $D5$-branes; but it is very difficult to see the effects of a RR background potential value on branes in the NS sector. 

It is however still possible to analyze the effect on the spectrum of strings, as we did in the previous Subsection, by realizing the discrete theta angle in the AOB background with a holonomy of $(-1)^{F_L} T$. This acts as
\begin{equation}\left(\begin{array}{c}C_2\\B_2\end{array}\right)\,\rightarrow\, \left(\begin{array}{c}-C_2-B_2\\B_2\end{array}\right),\end{equation} 
so now the field $B_2$ is invariant and the combination $B_2+2C_2$ flips sign. As a result, the 9d 2-form tensor corresponds exactly to the ordinary perturbative $B_2$ field (just as in the AOB background with no theta angle); an important consequence is that we still have a perturbative, BPS string in the spectrum. From \eq{e32co}, however, we can conclude that a $D1$ brane ($q=1$) has \emph{half} the charge under $B_2$ than a BPS string. Since the $D1$ is also charged under the combination that is projected out (and that becomes a $\mathbb{Z}_2$-valued field in the nine-dimensional theory), we conclude that, in the AOB background with discrete theta angle, \emph{there is a non-BPS string with half the fundamental string charge}. Just as for the DP example above, BPS completeness fails. The reason why a $D1$ ends up being charged under the NS-NS field is easy to understand in the covering space picture; a $D1$ is not invariant under $(-1)^{F_L} T$, and its image is a $(-1,1)$ string which is also charged under the $B_2$-field. The $\mathbb{Z}_2$ identification then divides charges out by two, effectively introducing the fractional charge.

Since the fundamental string is BPS and is not projected out, we can access the spectrum of BPS particles via a worldsheet description. The ordinary AOB background introduced in \cite{Hellerman:2005ja} constructs the worldsheet description by orbifolding the sigma model of IIB on $S^1$ by the symmetry which half-shifts the circle and acts by $(-1)^{F_L}$; what we need to do now is simply replace $(-1)^{F_L}$ by $(-1)^{F_L} T$. Since the  $T$ transformation acts nontrivially on the pair $(C_2,B_2)$, it must act non-trivially on the worldsheet too, on the RR 1-form and NS vertex operators. We will now use consistency of the spacetime picture to obtain the correct worldsheet CFT for the fundamental BPS string in this setup. Although this will be enough for our purposes, it would be an interesting follow-up to our work to explore this new string theory in the worldsheet directly,  as an orbifold of IIB on a circle.

In this note, we will focus our attention on two duality invariant pieces of information of the moduli space: the current algebra levels \cite{Kim:2019vuc,Kim:2019ths}. As described in \cite{Kim:2019ths}, in models of sixteen supercharges one expects an anomalous Bianchi identity for the 2-form in the gravity multiplet, which on an enhanced symmetry locus takes the form 
\begin{equation} dH= \ell\,\text{Tr}(F^2)-\kappa\, \text{Tr}(R^2).\label{ejgh}\end{equation}
The gauge level $\ell$ and the gravitational coefficient $\kappa$ are important topological data in specifying the supergravity theory, which can be determined from anomaly inflow on the string worldsheet. As we will see momentarily, we have $\kappa=0$, just as for the ordinary AOB background. 

The level $\ell$ would be apparent if we had the worldsheet description; here, instead, we will use a spacetime argument to compute $\ell$, and use it to determine the worldsheet uniquely. As described above, there are actual strings in our theory with charge $1/2$ of that of the fundamental string. By Dirac quantization, one must have then that the fundamental magnetic fourbrane charge is a multiple of $2$, meaning that its Dirac pairing with the fundamental BPS string is two. Indeed, this is what we derive from the microscopics: magnetic four-branes come from wrapping ten-dimensional IIB 5-branes on a circle. Fivebrane charges transform exactly as the $(C_2,B_2)$ fields in \eq{s2zbct}, and so single NS5 brane is not invariant under the action of $\Omega T$ and is projected out. The object of smallest magnetic charge under $B_2$ which is not projected out by the $\Omega T$ action is the $(2,-1)$ brane, which indeed has twice the charge of a single NS5-brane (and also happens to be BPS). 

As is familiar from higher-rank cases \cite{Font:2020rsk,Font:2021uyw}, in special loci in moduli space there may be an enhanced gauge symmetry, where the gauge group becomes non-abelian. In the case at hand, since the theory is of rank 1, the only possible enhancement is to something with $\mathfrak{su}(2)$ Lie algebra. 
Due to the coupling \eq{ejgh}, in a 9d $\mathcal{N}=1$ theory, the BPS instanton solutions of the low-energy supergravity, whose existence is unavoidable, acquires a four-brane magnetic charge of precisely $\ell$, the level of the gauge algebra. From the above argument, it is clear that $\ell$ must be an even number. On the other hand, the central charges of the worldsheet CFT must be $(c_L,c_R)=(12,12)$, since the worldsheet theory is obtained via an orbifold of the usual IIB on $S^1$ worldsheet, which has these central charges. This means that $\kappa=0$ as advertised above. In Section \ref{sec:comp} we will recover this result again, from consistency with the M theory picture.

Given that the worldsheet theory must preserve $SO(8,1)$ rotational invariance, must have at least 8 unbroken supercharges, and have an $\mathfrak{su}(2)$ current algebra at level two (at the self-dual point), the answer is essentially unique: The only possibility is to replace the $\mathcal{N}=1$ left-moving supermultiplet $(\phi,\lambda)$ describing the internal $S^1$ in the type IIB model by an $\mathfrak{su}(2)$ current algebra at level two. The central charge of this model is $3/2$, which exactly matches that of the removed supermultiplet, so that  the left-moving central charge is still 12.

The worldsheet model we propose here has manifest $(0,8)$ supersymmetry. It may be that additional supercharges are non-linearly realized in the left-moving sector; for instance, the WZW model at level two has emergent $\mathcal{N}=1$ supersymmetry \cite{Bae:2021lvk}, and this may combine with the supercharges in the center-of-mass modes to produce $(8,8)$ supersymmetry. It would be interesting to study this further\footnote{Reference \cite{Kim:2019ths} claims that only $(8,8)$-supersymmetric worldsheets are compatible with $\kappa=0$. The authors of \cite{Kim:2019ths} point out to the Asymmetric IIA and IIB orbifold backgrounds as examples. In the AOB model, for example, the worldvolume fields arrange themselves into $(8,8)$ multiplets, precisely the same sigma model as IIB on a circle, with an unusual GSO-like projection.}.

Finally, when its size modulus is sent to zero, the gauge instanton becomes a fundamental brane of the theory, which can often be identified with the branes constructed by microscopic means. For instance, in heterotic string theory, the small instanton limit corresponds to the NS5 brane \cite{Witten:1995gx}. It is natural to guess that, in the AOB background with discrete theta angle, the small instanton limit corresponds to the $(2,-1)$ IIB brane wrapped on a circle.

Just like the DP background above, the AOB background has many discrete fields, this time coming from the RR sector (the NS NS fields, the metric and $B$ field, yield the metric, 2-form, and two vectors of the 9d theory, and no discrete fields). The following table summarizes the discrete fields that exist in this theory, and the corresponding electrically \& magnetically charged objects (here, an $n$-brane has $n$ spatial dimensions, e.g. the 7-brane is a $(7+1)$-dimensional object)
 
  \begin{equation} \begin{array}{c|c|c}\text{$\mathbb{Z}_2$-valued field}&\text{Object}&\text{Stringy origin}\\\hline
  \multirow{2}{*}{$C_0$}&(-1)\text{-brane (electric)}&D(-1)\, \text{instanton}\\  &7\text{-brane (magnetic)}&\text{Unwrapped}\, 7\, \text{brane}\\\hline
   \multirow{2}{*}{$C_2$}&1\text{-brane (electric)}&D1\, \text{brane}\\  &5\text{-brane (magnetic)}&\text{Unwrapped}\, 5\, \text{brane}\\\hline
   C_4\,\text{(self-dual)}&3\text{-brane }&\text{Unwrapped}\, D3\, \text{brane}
  \end{array}\label{eq:w232}\end{equation}

 We conclude by remarking that the AOB background, just like the DP above, provides an example of a string with a nontrivial sublattice of charged BPS strings. Equivalently, it provides an example of sublattice WGC for strings. The fact that the example has sixteen supercharges means it can be analyzed quite detailedly, even at the non-perturbative level. As we will see momentarily, the similar properties of the DP and AOB backgrounds stem from the fact that the two are actually dual to each other.

 \subsection{Duality and the moduli space}\label{subsec:dual}
 
 Having described both the discrete theta angles in the DP and AOB backgrounds, one might wonder how they are related by duality, or whether there might be additional theta angles by looking at more general backgrounds. To do this, we will now study the most general background of IIB on a $S^1$ with a Wilson line that preserves some supersymmetry. This is just a IIB compactification on a circle with a duality bundle. The duality group of IIB is often presented as $SL(2,\mathbb{Z})$, but this is inaccurate; including $\Omega$ and $(-1)^{F_L}$ upgrades this to $GL(2,\mathbb{Z})$ at the bosonic level, and considering fermions further promotes this to a $\text{Pin}^+$ double cover of $GL(2,\mathbb{Z})$ \cite{Pantev:2016nze,Tachikawa:2018njr}. Fermions will not play an important role in our current discussion, so when we say ``duality group'' we will mean $GL(2,\mathbb{Z})$.   

Bundles of non-abelian groups such as $GL(2,\mathbb{Z})$ on a circle are specified by choosing a conjugacy class $[W]$; this implements the fact that both $W$ and $g\,Wg^{-1}$ for any $g$ in the duality group have the same physical effect. So we are led to studying conjugacy classes of $GL(2,\mathbb{Z})$ \cite{Debray:2021vob}. However, not every conjugacy class will yield a valid background; we must choose a vev for the axio-dilaton $\tau$, which is invariant (up to duality transformation) as we go around the circle. The action of a general $g\in GL(2,\mathbb{Z})$ on $\tau$ is as follows \cite{Pantev:2016nze,Tachikawa:2018njr}:
\begin{equation}\rho_g(\tau)=\frac{a\tilde{\tau}+b}{c\tilde{\tau}+d},\quad\text{where}\quad \tilde{\tau}=\begin{cases}\tau&\text{if}\, \text{det}(g)=+1\\ \bar{\tau}&\text{if}\, \text{det}(g)=-1
\end{cases}\quad .\end{equation}
We must choose $[W]$ such that the equation
\begin{equation}\tau=\rho_{g W g^{-1}}(\tau)\label{ew2}\end{equation}
has a solution in the upper half-plane. Working this out is a standard algebraic exercise \cite{bullett2017geometry}. Write
\begin{equation} h=g W g^{-1}=\left(\begin{array}{cc}a&b\\c&d\end{array}\right).\end{equation}
When $\det h=+1$, so that we are in $SL(2,\mathbb{Z})$, \eq{ew2} becomes a quadratic equation, with solution
\begin{equation}\tau=\frac{a-d\pm\sqrt{(d+a)^2-4}}{2c}.\end{equation}
The requirement that the solution is in the upper half-plane leads to the condition $\vert d+a\vert <2$, which only has solutions 
\begin{equation}\left(\begin{array}{cc}a&b\\c&d\end{array}\right)=\pm\left\{\left(\begin{array}{cc}0&-1\\1&0\end{array}\right),\, \left(\begin{array}{cc}-1&1\\-1&0\end{array}\right),\, \left(\begin{array}{cc}0&-1\\1&-1\end{array}\right)\right\}.\end{equation}
We recognize the $S$, $U$ and $U^{-1}$ elements of $SL(2,\mathbb{Z})$. Only for these conjugacy classes it is possible to compactify on a circle with duality Wilson line. This has been recently used in lower-dimensional compactifications to produce new IIB backgrounds and associated dual CFT's \cite{Assel:2018vtq,Bobev:2021yya,Guarino:2021hrc,Cesaro:2021tna,Cesaro:2022mbu}, but in nine dimensions these Wilson lines do not preserve any supercharges \cite{Tachikawa:2018njr,Debray:2021vob}. Therefore, we now move to Wilson lines with $\det h=-1$. In this case, the equation to solve is
\begin{equation}\bar{\tau}=\frac{a\tau+b}{c\tau+d},\end{equation}
which becomes a system of two equations for $\tau= x+i\,y$,
\begin{equation}cy^2=-cx^2+(a-d)x+b,\quad (d+a)y=0.\label{bilet}\end{equation}
The second equation and the condition $\det h=-1$ together imply
\begin{equation} a=-d,\quad d^2+bc=1.\end{equation}
These equations have infinitely many solutions. Here, we only consider the following simple solutions, leaving the rest for Appendix \ref{appA}:
\begin{equation}\left(\begin{array}{cc}a&b\\c&d\end{array}\right)=\pm\left\{\left(\begin{array}{cc}0&1\\1&0\end{array}\right),\, \left(\begin{array}{cc}1&b\\0&-1\end{array}\right),\, \left(\begin{array}{cc}1&0\\c&-1\end{array}\right)\right\}.\label{w2344}\end{equation}
The first entry above corresponds to a holonomy in the conjugacy class $[\pm S\Omega]$ and, just like the possibilities above, it does not preserve any supercharges.

 The second entry, corresponding to $\Omega\, T^b$ and $(-1)^{F_L}\, T^b$, precisely captures the DP and AOB backgrounds, with or without discrete theta angle depending on whether $b$ is even or odd. Actually, due to the fact that 
 \begin{equation}\Omega\, T^{b}=T^{-b}\Omega\end{equation}
  in $GL(2,\mathbb{Z})$, all values of $b$ differing by an even number are in the same conjugacy class: 
 \begin{equation}T^n\Omega\, T^{b}\, T^{-n}=\Omega\, T^{b-2n}.\end{equation}
 Hence only $b=0,1$ are relevant. $b=0$ is the case with no theta angle, and $b=1$ corresponds to theta angle turned on.  It will be important below that $[\Omega\, T^b]$ and $[(-1)^{F_L} T^b]=[-\Omega\, T^b]$ are actually the same conjugacy class, as can be seen by conjugating by $ S T^{-2}S^{-1}$:
 \begin{equation} (S T^{-2}S^{-1})\, \Omega T^b\,( S T^{-2}S^{-1})^{-1}= T^b\, (-1)^{F_L}.\label{wq2}\end{equation}
  The third entry in \eq{w2344} describes a Wilson line in the conjugacy class of $[S\, \Omega\, T^c\,S^{-1}]$ or $[S\, (-1)^{F_L}\, T^c\,S^{-1}]$. Just as before, only $c$ mod 2 is physically meaningful. These are the same conjugacy classes as for the DP and AOB backgrounds, but described in a dual frame. 
 
 To sum up, a systematic analysis reproduces the backgrounds we discussed already, and nothing else; therefore, the theta angles that we have discussed exhaust all the possibilities in 9d coming from circle compactifications of type IIB with a duality Wilson line.  We wrap up by describing S-duality for these backgrounds. The S-transformation preserves the line $\tau=i/g_s$, sending $g_s\rightarrow 1/g_s$, so when the discrete theta angle is turned off, it implements a strong-weak coupling duality, allowing one to describe the regime $g_s\gg 1$ in terms of a dual, weakly coupled string of the same kind. When the discrete theta angle is turned on, the line $\tau=i/g_s-1/2$ is not invariant under the S-transformation, and in particular it is mapped to the line
 \begin{equation}\tau=-\frac{1}{1/2+i/g_s}=\frac{4g_s}{4+g_s^2}\left[-\frac12+\frac{i}{g_s}\right].\end{equation}
 Thus, the S-transformation maps weak coupling to weak coupling, and provides no interesting information at strong coupling. As suggested by \eq{wq2}, the correct duality transformation involves the element $V\equiv S T^{-2}S^{-1}$, which maps $\tau$ as
 \begin{equation} \tau=-\frac12+\frac{i}{g_s}\,\rightarrow\frac{\tau}{1+2\tau}=\frac12+\frac{g_s}{4} i \sim-\frac12+\frac{g_s}{4} i  ,\end{equation}
 where in the last equality we have used that $\tau\sim \tau+1$ via a T transformation, and so it relates the weak coupling ($g_s\ll1$) DP background to the strong coupling ($g_s\gg1$) AOB background and vice-versa, just like the ordinary S-transformation does when there is no theta angle. The fixed point is $g_s=2$, as opposed to $g_s=1$ in the usual $S$-duality. This transformation shows that the discrete theta angle affects the details, but not the qualitative nature, of strong-weak duality. There are, however, some unusual features. The usual S-duality exchanges fundamental strings and D1 branes. This can be seen from the action of the transformation on the field doublet (we use the conventions of \cite{Weigand:2018rez})
 \begin{equation}\left(\begin{array}{c}C_2\\B_2\end{array}\right)\,\rightarrow\, \left(\begin{array}{cc}0&-1\\1&0\end{array}\right)\left(\begin{array}{c}C_2\\B_2\end{array}\right)=\left(\begin{array}{c}-B_2\\C_2\end{array}\right).\label{s2zbct}\end{equation} 
By contrast, the action of the transformation $V=S T^{-2}S^{-1}$ maps
 \begin{equation}\left(\begin{array}{c}C_2\\B_2\end{array}\right)\,\rightarrow\, \left(\begin{array}{cc}1&0\\2&1\end{array}\right)\left(\begin{array}{c}C_2\\B_2\end{array}\right)=\left(\begin{array}{c}C_2\\2C_2+B_2\end{array}\right),\label{ew222}\end{equation}
 so the $C_2$ field is mapped to itself, and $B_2$ is shifted. This means that charged objects mix with each other. For concreteness, consider the DP background at weak coupling. The fundamental string is $\mathbb{Z}_2$-charged, and the D1 string has a $\mathbb{Z}$-valued charge and is a BPS object. in the dual AOB background, the fundamental string is BPS and has a $\mathbb{Z}$-valued charge, while the $D1$ brane is $\mathbb{Z}_2$-charged. The transformation \eq{ew222} acts via the transpose transformation on the charges, and it implies that 
 \begin{equation} \text{ Fundamental string of DP}=\text{ Fundamental AOB string} - 2 D1\text{'s of AOB}.\end{equation}
The tensions of these two objects match, as they should. As mentioned in Subsection \ref{subsec:10dsethi}, in Planck units, the tension of a $(p,q)$ string is \cite{polchinski1998string}
 \begin{equation} T_{(p,q)}=\frac{\vert p+q\tau\vert}{\sqrt{\text{Im}(\tau)}}.\end{equation}
 One can see that both $T_{(1,0)}$ for $\tau=i/g_s-1/2$ and $T_{(-2,1)}$ for $\tau=\frac12+\frac{g_s}{4} i$ agree and are equal to $\sqrt{g_s}$. The ratio of tensions between the dual fundamental string and the original fundamental string is $2/g_s$, a factor of 2 larger than for the usual IIB S-duality \cite{polchinski1998string}. This factor of two is precisely the index of the sublattice of BPS  that we found in the previous Subsection; in this background, the perturbative fundamental string has charge twice the fundamental charge. We also notice that the $\mathbb{Z}_2$-valued fields that we described in this subsection and the previous one are perfectly matched to each other under S-duality; RR fields in DP map to RR discrete fields in AOB, and vice-versa. It is interesting that RR fields are captured by the K-theory description in both cases, while there is no such description available for the NS-NS fields.
 
 The duality group of this theory is a subgroup of $SL(2,\mathbb{Z})$ generated by 
 \begin{equation} V=\left(\begin{array}{cc}1&0\\2&1\end{array}\right)\quad\text{and}\quad T=\left(\begin{array}{cc}1&1\\0&1\end{array}\right). \end{equation}
 These generate the Hecke congruence subgroup $\Gamma_1(2)$, defined as the subgroup of $SL(2,\mathbb{Z})$ composed of matrices satisfying
 \begin{equation} M=\left(\begin{array}{cc}a&b\\c&d\end{array}\right),\quad a\equiv d\equiv 1\,\text{mod}\,2,\quad c\equiv 0\,\text{mod}\,2.\end{equation}
 The fact that $\Gamma_1(2)$ appears as the duality group of an explicit string construction is sure to have further implications. For instance, toroidal compactifications of the AOB background to four dimensions will produce $\mathcal{N}=4$ models where the duality group is $\Gamma_1(2)$ rather than the usual $SL(2,\mathbb{Z})$. It would be interesting to study these in detail.  The fact that the duality group is a congruence subgroup of $SL(2,\mathbb{Z})$ is in line with the expectation put forth in \cite{Dierigl:2020lai}, and is related to the fundamental group of the moduli space being purely torsion as required by Swampland principles \cite{Ooguri:2006in}.

\subsection{The moduli space of rank 1 nine-dimensional compactifications}
 
Armed with this new version of S-duality, we can now understand all the corners of the new component of the moduli space we found. Just like any other 9d $\mathcal{N}=1$ rank 1 quantum gravity, the theories we discussed in the previous Subsections have a two-dimensional moduli space, parametrized by the dilaton (a scalar in the gravity multiplet) and a scalar from the vector multiplet (the size of the $S^1$ for the AOB or DP backgrounds). The geometry of the moduli space is purely determined by supergravity, and thus in particular it is insensitive to the presence of discrete theta angles; however, as we will see, the duality webs are significantly different.

\begin{figure}[hbtp!]
\centering 
    \includegraphics[width=0.8\textwidth]{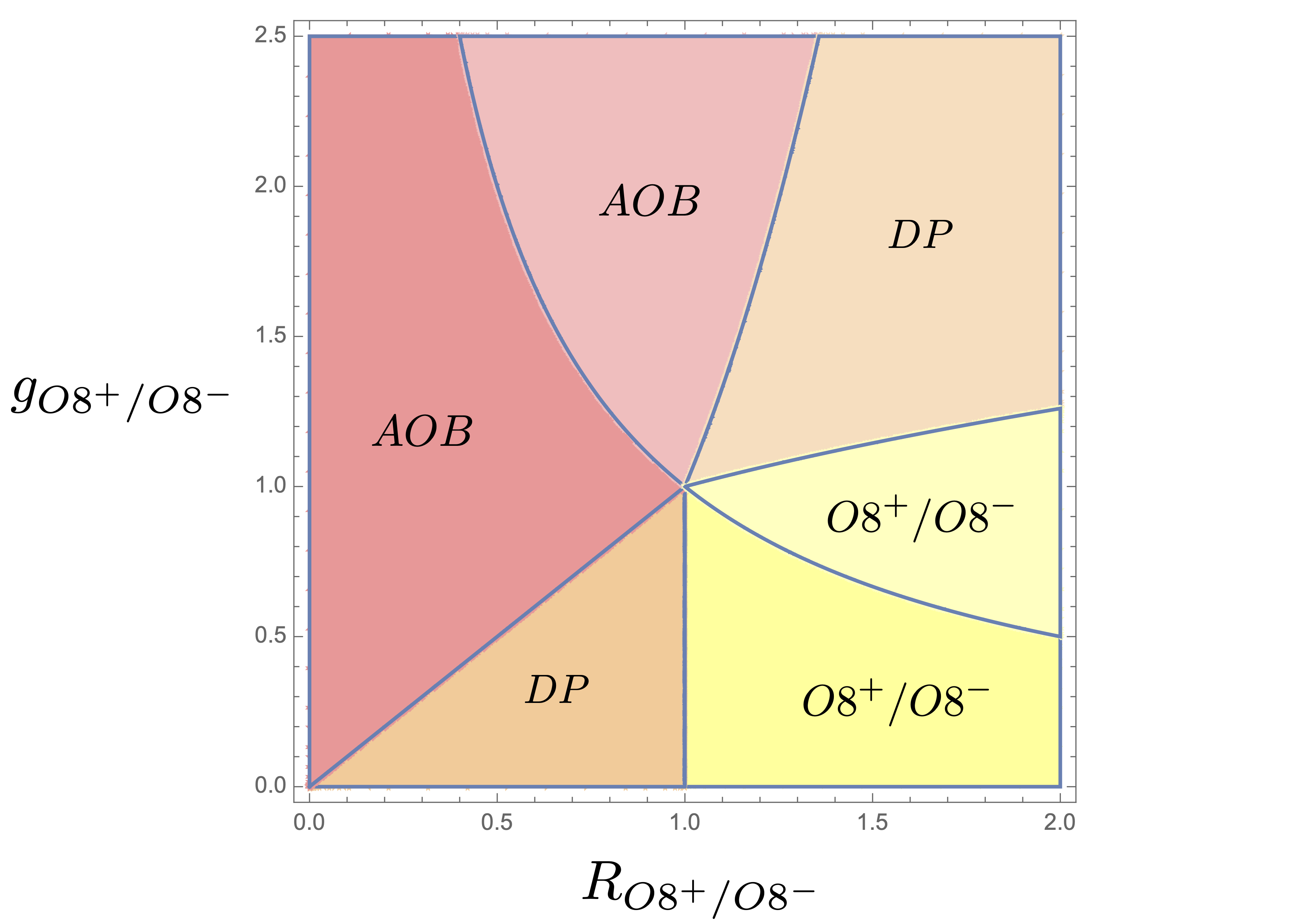}
    \caption{Depiction of the moduli space of the rank 1 component of the nine-dimensional moduli space including the Dabholkar Park, AOB, and $O8^+/O8^-$ compactifications. Following \cite{Aharony:2007du}, from which this picture is taken, we take the $O8^+/O8^-$ component as reference and parametrize the whole moduli space in terms of its coupling and distance between O-planes. For each value of these parameters the figure indicates which description is weakly coupled. The curved line running from top left to bottom right of the picture is a self-duality line, and points to both sides of the line are to be identified; this is also encoded in the color, where regions with different shades of the same color are to be identified.}
    \label{f1}
\end{figure}

Let us first describe the moduli space with theta angle turned off. The moduli space of the DP background is carefully explored in the beautiful paper \cite{Aharony:2007du}; by moving on the dilaton/radius space, one can reach other perturbative corners, admitting a dual description. In particular,  when the discrete theta angle is turned off, one reaches other two perturbative descriptions, in different limits in moduli space:
\begin{itemize} 
\item At strong coupling, one S-dualizes to IIB on $S^1$ with a Wilson line of $(-1)^{F_L}$, i.e. the AOB background in \cite{Hellerman:2005ja,Aharony:2007du}, with vanishing theta angle. 
\item At small radius, a T-dual IIA description emerges, in terms of an interval compactification with an $O8^+/O8^-$ orientifold pair and no branes.
\end{itemize}
On the AOB corner, there is a locus of enhanced $SU(2)$ symmetry, at a self-dual value of the radius. This is also visible in the $O8^+/O8^-$ description at strong coupling, where increasing the coupling at the $O8^-$ can cause $D0$ branes stuck there to become massless, providing the required enhancement (see the discussion in Subsection \ref{sec:sethi}). This state of affairs is conveniently depicted in Figure \ref{f1}, which we took  from \cite{Aharony:2007du}. As usual, the T-duality line is actually an identification on the physical moduli space of the theory, with theories at either side of the line being physically equivalent. The figure thus depicts a double cover of the actual moduli space, which is like a napkin folded over itself. 

Let us now explain how the above description is modified in the presence of the discrete theta angle. As described above, both the AOB and the DP  backgrounds admit their own versions of discrete theta angles, but they are not the only ones to do so. For the $O8^+/O8^-$ background, the theta angle is a version of the idea in \cite{Sethi:2013hra} that we described in Section \ref{sec:sethi}; the orientifolds project out the RR field $C_1$, but leave out the possibility of having a Wilson line
\begin{equation} \int C_1=1/2\end{equation}
on the covering circle. This means that, as a $D0$ brane moves from the $O8^-$ to the $O8^+$ and back again, its wavefunction picks a factor of $(-1)$. In Subsection \ref{subsec:10dsethi} we tried to do this in compactifications involving $O8$ planes and $D8$ branes; as explained there (and following \cite{Witten:1998cd,Bergman:2013ala}), the $D8$ branes make the $\int C_1$ angle unphysical, since it is equivalent to moving to a different point in moduli space. But there are no $D8$ branes in the rank 1 case, so the argument does not apply and the angle is physical this time.

This $O8^+/O8^-$ theta angle is just the T-dual of the DP one in Subsection \ref{subsec:aobth}, since T-duality turns $C_0\leftrightarrow C_1$. So the moduli space of the DP background with theta angle also includes a T-dual corner with a $O8^+/O8^-$ theta angle. One advantage of the IIA description is that it now becomes possible to access the point of enhanced symmetry. As described in Section \ref{sec:sethi}, when a discrete Wilson line $\int C_1$ is turned on, an $O8^-$ does not receive massless degrees of freedom coming from massless $D0$ branes; but instead it is able to non-perturbatively emit an additional $D8$. This $D8$ can move all the way to the other side of the interval, at the $O8^+$ plane, to realize an $\text{Sp}(1)\sim SU(2)$ enhanced symmetry\footnote{We note in passing that the existence of this theory was actually implied by the Swampland arguments in \cite{Bedroya:2021fbu}, although we did not realize it at the time. A rank 1 theory has a duality group given by $O(1,1,\mathbb{Z})=\mathbb{Z}_2$, so there can be at most one enhanced symmetry locus per moduli space component. In Table 4 of \cite{Bedroya:2021fbu}, all possible maximal symmetry enhancements of rank 1 theories are listed, and there are three of them. Thus, the classification predicted one more moduli space component than the two that were known at the time.}. This happens at a locus $g\sim \frac{2}{R}$, which therefore marks the regime of validity of the type I' description. As is the case for the similar symplectic factors that appear in the CHL string, the $\text{Sp}(1)$ current algebra is at level 2, just like the AOB background described in Section \ref{subsec:aobth}. 

\begin{figure}[hbtp!]
\centering 
    \includegraphics[width=0.8\textwidth]{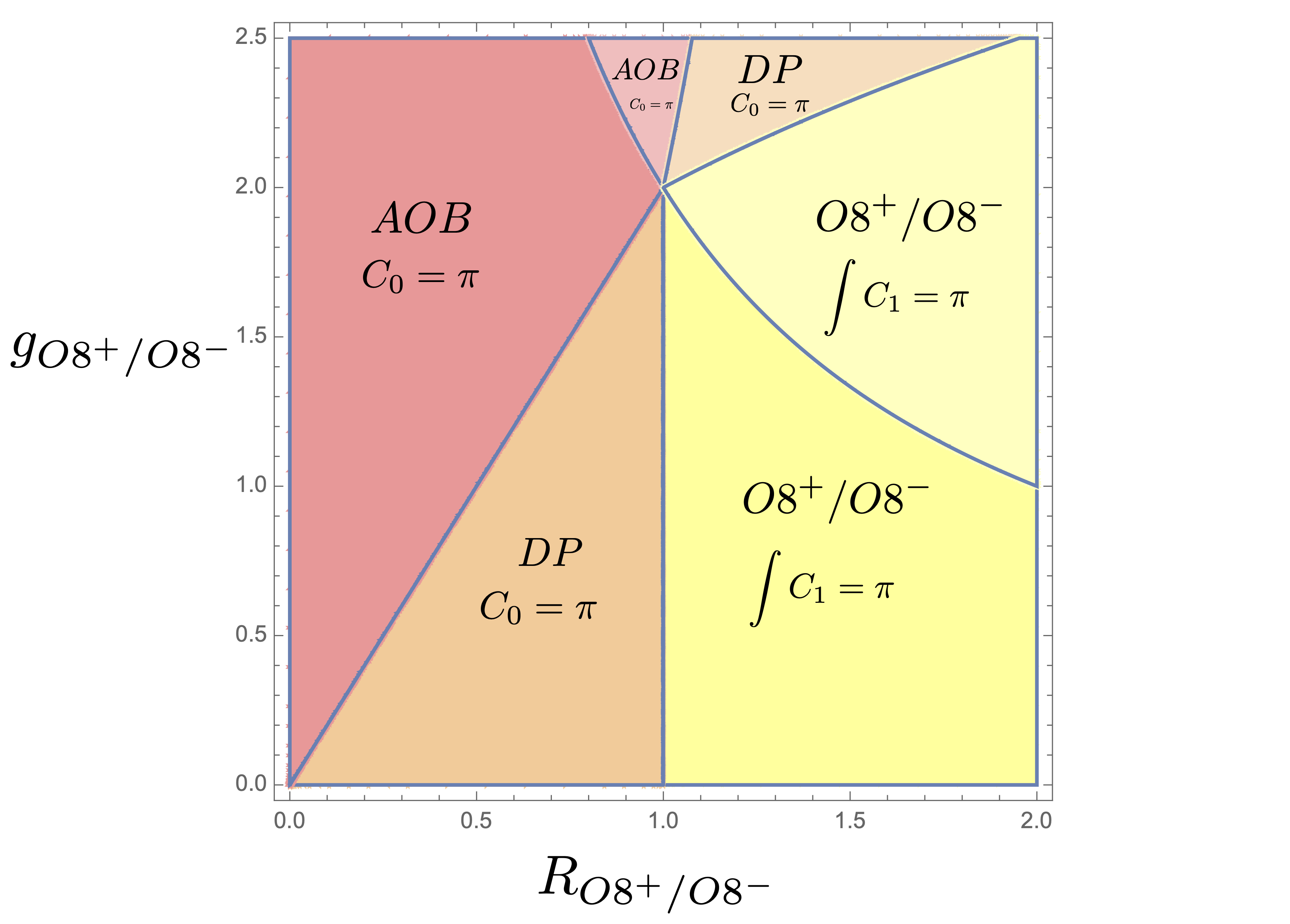}
    \caption{Depiction of the new component of the nine-dimensional rank one moduli space. As described in the main text, it is fully covered by versions of the Dabholkar Park, AOB, and $O8^+/O8^-$ compactifications with discrete theta angles turned on. As in Figure \ref{f1} we used the $O8^+/O8^-$ component to parametrize all of moduli space. The only difference with Figure \ref{f1} is the scale of the vertical axis, reflecting the rescaling \eq{resc5}. Relatedly, the self-dual point is at coupling $g_s=2$, and the duality group is $\Gamma_1(2)$ instead of the $SL(2,\mathbb{Z})$ that one obtains when the discrete theta angle is switched off.}
    \label{f1b}
\end{figure}

We are now ready to show that all the new theories described thus far lie in the same moduli space, and that cover it completely. The argument is essentially the same as that of \cite{Aharony:2007du} for the component without discrete theta angle, with just a few additional factors of two. We parametrize the whole moduli space by the coupling and interval size $(g_\pm, R_\pm)$ of the $O8^+-O8^-$ theory with discrete theta angle turned on; the range of validity of this corner of moduli space is given by $R_\pm\gtrsim 1$ and $g_{\pm}<2/R_\pm$. Decreasing $R_\pm$ at constant $g_\pm$ forces us into a T-dual DP background, with $C_0$ theta angle turned on, and T-dual couplings related to the original ones by \cite{polchinski2005string}
\begin{equation} g_{\text{DP}}=\frac{g_\pm}{R_\pm},\quad R_{\text{DP}}=\frac{1}{R_{\pm}}.\end{equation}
The regime of validity of this new T-dual background is set by the self-dual line $g_{\text{DP}}=2$, $R_{\text{DP}}>1$, or 
\begin{equation} g_{\pm}\leq 2R_{\pm},\quad R_{\pm}\leq 1.\end{equation}
Decreasing $R_{\pm}$ even further, past the strong coupling line, forces us to perform $S$-duality as explained in Subsection \ref{subsec:dual}. The S-dual AOB description has couplings
\begin{equation}g_{\text{AOB}}=\frac{4}{g_{\text{DP}}}=\frac{4R_\pm}{g_\pm},\quad R_{\text{AOB}}=R_{\text{DP}}\sqrt{\frac{\alpha'_{\text{DP}}}{\alpha'_{\text{AOB}}}}=\sqrt{\frac{2}{R_\pm g_\pm}},\end{equation}
where the quotient between the string length of the fundamental DP and AOB strings is
\begin{equation}\frac{\alpha'_{\text{DP}}}{\alpha'_{\text{AOB}}}=\frac{T_{(1,0),\text{AOB}}}{T_{(1,0),\text{DP}}}=\frac{2}{g_s}\end{equation}
as derived in Subsection \ref{subsec:10dsethi}. We expect the self T-duality line at $R_{\text{AOB}}=1$, or equivalently, at
\begin{equation} R_\pm g_\pm\sim 2\end{equation}
The T-dual background has couplings
\begin{equation} g_{\text{AOB-II}}=\sqrt{\frac{8R^3}{g}},\quad R_{\text{AOB-II}}=\sqrt{\frac{gR}{2}}.\end{equation}
Now increasing $R_\pm$ again, we hit an S- dual $DP$ background, with couplings
\begin{equation} g_{\text{DP-II}}=\sqrt{\frac{2g}{R^3}},\quad R_{\text{DP-II}}=\left(\frac{g^3}{8R}\right)^{\frac14}.\end{equation}
Finally, increasing $R_\pm$ even further leads us back to an $O8^+-O8^-$ background, with couplings
\begin{equation} g_{\pm\text{-II}}=\left(\frac{32}{gR^5}\right)^{\frac14},\quad R_{\pm\text{-II}}=\left(\frac{8R}{g^3}\right)^{\frac14}.\end{equation}
Importantly, and just like in the case with no theta angle, the lines
\begin{equation} g_{\pm\text{-II}} R_{\pm\text{-II}}=2\quad\text{and}\quad g_\pm R_\pm=2\end{equation}
coincide. This means that we have derived, indirectly, the strong coupling limit of the $O8^+-O8^-$ compactification to be itself\footnote{Note that this is an example of a compactification with Romans' mass turned on, for which the strong coupling limit exists and is known exactly. However, as we send $g_\pm$ to a large value, gradients and curvatures in the interval grow without bound. This is consistent with the results of \cite{Aharony:2010af}, which establishes that there is no strong coupling limit of a massive IIA configuration with low curvatures. The results of \cite{Aharony:2007du} that we have reviewed here show that high-curvature, strongly coupled limits of massive IIA actually exist.}. All six regions thus obtained completely cover a copy of the $SO(1,1,\mathbb{R})$ moduli space that corresponds to a rank 1 theory. In fact, the regions we have obtained are exactly the same as one obtains for the component of moduli space with theta angle turned off, which is depicted on Figure \ref{f1}, subject to the rescaling
\begin{equation} g_\pm\rightarrow \frac{g_\pm}{2} \label{resc5}\end{equation}
The resulting diagram is depicted in Figure \ref{f1b}.
This shows that the moduli space picture is completely consistent, and that all the new string compactifications discovered so far (discrete theta angles in DP and AOB, as well as in $O8^+-O8^-$) are all corners of the same, new component of the moduli space of 9d $\mathcal{N}=1$ string compactifications.

\section{Discrete theta angles in 8d string compactifications with sixteen supercharges}\label{sec:comp}
The next natural step is to look for theta angles leading to new string theories with 16 supercharges in 8 dimensions. Just as before, we will begin with a short review of the state of the art. Previous literature only discusses a single component of the rank 2 moduli space in eight dimensions \cite{Aharony:2007du}; it has two different decompactification limits, leading to the two previously known rank 1 components of the moduli space in nine dimensions. This unique component of the moduli space has several corners, which are conveniently described as 
\begin{itemize} 
\item The $O8^+-O8^-$ on a circle;
\item After T-dualizing the additional circle, the $O8^+$ becomes a pair of $O7^+$'s and similarly for the $O8^-$; the configuration becomes a IIB compactification on $T^2/\mathbb{Z}_2$ with two $O7^+$ planes and two $O7^-$ planes, which can be equivalently described as F-theory on K3 with two frozen singularities \cite{Aharony:2007du}.
\item We can now T-dualize on the circle on which the $O7^+/O7^-$ pairs extend. This way we reach a new decompactification limit, where the $O7^+$ and $O7^-$'s pair up to produce a pair of two $O8^0$'s; the configuration we are describing is the compactification of the Asymmetric IIA Orbifold, or AOA background \cite{Aharony:2007du}, on a circle. This background admits a further uplift to M-theory on a Klein bottle \cite{Hellerman:2005ja}.
\end{itemize}

Let us now look for new string theories. The new component of moduli space we found in nine dimensions immediately produces, when compactified on a circle, a new component of the moduli space in 8 dimensions. This can be described as $O8^+-O8^-$  with $\int C_1=1/2$ on a circle or, after T-duality,
\begin{equation*} \text{IIB on an $O7^+-O7^-$ background with}\quad \int_{T^2}C_2=1/2.\label{msug}\end{equation*}
The story is by now familiar. Consider modding out IIB on $T^2$ by the symmetry that flips both coordinates of the $T^2$ and acts simultaneously with an action of $\Omega$. This projects out several fields, notably the holonomy $\int_{T^2}C_2$. But a $\mathbb{Z}_2$ subgroup survives, providing a discrete theta angle. 

In \cite{Berglund:1998va}, it was proposed that a profile like \eq{msug} was the F-theory description of the CHL string. We believe it is rather more natural to just take F-theory on Möbius$\times S^1$ to be the F-theory description of the CHL string; it matches the usual M-F dictionary, and it follows the lore that we are supposed to be able to consider F-theory on any space with a $T^2$ fibration. The Mobius strip does not have such a fibration, but Mobius$\times S^1$  does. Be it as it may, it would be interesting to explore the connection with the picture in \cite{Berglund:1998va} in more detail. 

In nine dimensions, we could access two different corners of moduli space by choosing to act with $\Omega$ or $(-1)^{F_L}$; the same is true here, with the difference that only one of the descriptions is perturbative. The 7-branes that appear after orbifolding do not cancel their 7-brane charge locally, and so as usual in F-theory there is a logarithmic running of the 10d axio-dilaton near its core \cite{Weigand:2018rez}. Using $\Omega$ when defining the quotient will produce $O7$ planes at the fixed loci of the action on $T^2$, which are perturbative; the dilaton runs to weak coupling at their core. On the other hand, employing $(-1)^{F_L}$ will result in a non-perturbative configuration involving the $S$-duals of the $O7$ planes. This strongly coupled prescription is pretty much useless, and it is a good idea to S-dualize to the first case we described. In F-theory language, this follows from the statement that the perturbative limit (Sen's limit) is essentially unique; this is still true when frozen singularities are involved. The case without discrete theta angle corresponds to two frozen $D_8$ singularities together with a frozen torsional section of order two intersecting them \cite{Hector-paper}; switching on the theta angle gives a configuration without such a frozen section.    

As described above, when the theta angle is switched off, it is possible to $T$-dualize to a IIA description, given by an $O8^0$ compactification on a circle. Subsequently taking the strong coupling limit the interval containing the $O8^0$ planes becomes a Klein bottle in M theory, and so we end up with M-theory on Klein bottle times $S^1$. How does this change when the theta angle is turned on? Since the theta angle is a holonomy of $C_1$, which corresponds to pure geometry, we also expect to have an M-theory lift. One first guess could be that M-theory on KB$\times S^1$ admits a discrete theta angle coming from holonomy of the $C_3$ field, but this is not correct. The M-theory 3-form $C_3$ picks up an additional minus sign under the M-theory parity action \cite{Witten:1996md}, and so the period 
\begin{equation}\int_{\text{KB}\times S^1}C_3\end{equation}
is an ordinary circle-valued axion (the axion in the gravity multiplet of the eight-dimensional theory). The component of the three-form not along the Klein-bottle becomes a discrete $\mathbb{Z}_2$-valued 3-form field, which has featured prominently in the recent beautiful paper \cite{Yonekura:2016wuc}, where it was shown its presence is essential for the cancellation of a gravitational anomaly of Dai-Freed type in the nine-dimensional theory. However, a discrete 3-form in nine dimensions will not produce a discrete $\mathbb{Z}_2$ field in eight. So what is the origin of the eight-dimensional theta angle?

The answer turns out to be pure geometry. Consider a compactification of M-theory on the Klein bottle, described as the plane $\mathbb{R}^2$ with the Euclidean metric, and coordinates $(x,y)$ under the following identifications:
\begin{equation}\label{KBdef} (x,y)\,\sim\, (x+1,y)\,\sim(x,y+\tau_2)\,\sim\, (x+1/2,-y),\quad \tau_2\in\mathbb{R}.\end{equation}
The above space has a discrete $\mathbb{Z}_2$ isometry $\iota_1$, given by
\begin{equation} \iota_1:\, (x,y)\,\rightarrow\, (x,y+1/2).\end{equation}
There is also a second $\mathbb{Z}_2$ isometry $\iota_2$, given by
\begin{equation} \iota_2:\,(x,y)\,\rightarrow\, (-x,-y).\end{equation}
Each of these isometries give rise to exact discrete gauge symmetries of the nine-dimensional $\mathcal{N}=1$ gravity theory, and to the corresponding $\mathbb{Z}_2$-valued 1-form gauge fields. It is therefore possible to consider a compactification of the nine-dimensional theory on a circle with Wilson line for either $\iota_1,\iota_2$, or both. However, $\iota_2$ acts on the nine-dimensional supercharge as multiplication by $-1$, as we will show momentarily; therefore, the compactification with this Wilson line is non-supersymmetric\footnote{It can be analyzed from the worldsheet point of view, as a shift orbifold of a circle compactification of the AOA background; it would be interesting to pursue this analysis in detail, see if the compactification has tachyons, etc.}. To see this, consider the spinor lift of the last action in \eq{KBdef}. On two-dimensional spinors, it acts as 
\begin{equation}\psi(x,y)\,\rightarrow \Gamma_2\psi(x+1/2,-y),\label{fermact}\end{equation}
where $\Gamma_2$ is an Euclidean $\Gamma$ matrix which squares to $+1$. When one considers theories involving fermions and reflections, one must choose the action of the reflection on the fermions; there are two possibilities, depending on whether reflections square to $+1$ or to $-1$. See \cite{Witten:2015aba} for a nice exposition. The choice in \eq{fermact} corresponds to reflections squaring to $+1$, commonly called a Pin$^+$ structure; this is the symmetry type of M-theory \cite{Witten:1996md,Witten:2016cio}, as well as being the only action of reflections compatible with supersymmetry in nine dimensions \cite{Montero:2020icj}.

On the other hand, the spin lift of the action in $\iota_2$ is 
\begin{equation}\psi(x,y)\,\rightarrow  \Gamma_1\Gamma_2\psi(-x,-y),\label{fermact2}\end{equation}
corresponding to a rotation by $180^\circ$. Compactifying on $S^1$ with a Wilson line for $\iota_2$ means that spinors of the eight-dimensional theory must be invariant under both \eq{fermact} and \eq{fermact2}; there are no solutions to these equations at the level of fermion zero modes, since $\Gamma_1\Gamma_2$ and $\Gamma_1$ anticommute. Thus, we have shown that $\iota_2$ does not preserve spinors in eight dimensions. On the other hand, $\iota_1$ is just a translation, and it does not project out the supercharges; the resulting compactification is supersymmetric, and describes the discrete theta angle we found in other corners of the eight-dimensional moduli space. 

Since the $\mathbb{Z}_2$ symmetry $\iota_1$ we used in the construction has a geometric origin, we can directly describe the background M-theory is compactified on to produce this component of moduli space. The manifold is simply a mapping torus for the isometry $\iota_1$. Calling the coordinate for the additional circle as $z$, the manifold is fully specified as a quotient of $\mathbb{R}^3$ with coordinates $(x,y,z)$ subject to the identifications in \eq{KBdef} together with
\begin{equation} (x,y,z)\,\sim\, (x,y+1/2,z+1/2),\quad (x,y,z)\,\sim\,(x,y,z+1).\end{equation}
The resulting manifold is the quotient of $T^3$ by a freely acting isometry, and is automatically Ricci-flat. The most general such quotient is called a  Bieberbach manifold, and low-dimensional ones have been classified in the mathematical literature \cite{2013arXiv1306.6613L}. For instance, there are only six orientable Bieberbach three-dimensional manifolds, other than $T^3$. These appeared recently in \cite{Acharya:2019mcu}, where Acharya analyzed the possible spin structures on each of them. Although this six Bieberbach manifolds are Ricci-flat, and therefore solve Einsteins equations, none of them admit covariantly constant spinors. Therefore, they constitute interesting examples of classically stable solutions of Einsteins equations. They are not stable quantum-mechanically, either at large or small volume, as studied in \cite{GarciaEtxebarria:2020xsr,Acharya:2020hsc}.

Although reference \cite{Acharya:2019mcu} only looked at orientable Bieberbach manifolds, M-theory makes sense also in non-orientable manifolds \cite{Witten:1996md}. Non-orientable three-dimensional Bieberbach manifolds have also been classified; see Table 8 of \cite{2013arXiv1306.6613L}. There are four possibilities: $\text{KB}\times S^1$, and mapping tori of the Klein bottle for either $\iota_1,\iota_2$, and their product.  So the mathematical classification reproduces the backgrounds we found, and none else. As we showed above, not only $\text{KB}\times S^1$ admits covariantly constant pinors; the mapping torus by $\iota_1$, called $N_2^3$ in \cite{2013arXiv1306.6613L}, also does.  From this point of view, the new theory we describe in this paper is extremely simple: It just a compactification of M-theory in a non-orientable manifold which admits covariantly constant spinors. It would have turned up in a systematic construction of supersymmetric M-theory backgrounds, which has not been carried out even for the case of 16 supercharges that we are presently discussing.

Thus, to summarize, there is a new supersymmetric string theory in eight dimensions, which lives in a new component of moduli space. This component has two decompactification limits; one of them is the new component of the nine-dimensional moduli space we described in Section \ref{sec:new9d}, and the other one is a decompactification limit to M-theory on the Klein bottle, and is described as compactification on the background $N^3_2$. The classification in \cite{2013arXiv1306.6613L} shows that there are no further components of moduli space that can be accessed from the M-theory perspective.  

\section{Discrete theta angles in seven-dimensional theories}\label{sec:new7d}
We continue our journey by asking which new components of moduli space can be described in seven dimensions, the lowest number of dimensions in which sixteen supercharges correspond to minimal supersymmetry. As before, we start by considering the component of moduli space obtained from the new component we found in 8d and 9d via circle compactification. We will do this by considering the description involving $O7^\pm$ planes described in Section \ref{sec:comp}. Without theta angle, the resulting compactification was described in \cite{deBoer:2001wca}; one can T-dualize along the circle direction, and the $O7^\pm$ planes turn into $O6^\pm$ planes. Thus we have a three-dimensional IIA orientifold, without additional $D6$ branes.

How does the discrete theta angle affect this picture? We will not provide a proof, but we have sufficient information to make an educated guess. Recall that, in this duality frame, the discrete theta angle becomes an holonomy for the $C_2$ RR field on the covering torus. The worldsheet description is, in principle, insensitive to the theta angle; so the basic rules of T-duality should still apply. It follows that, after T-duality, one ends up with a $\mathbb{T}^3/\mathbb{Z}_2$ configuration with no branes, and equal numbers of $O6^+$ and $O6^-$ planes. However, the worldsheet does not have any way to access which of these are $O6^+$ and which are $O6^-$, since there are no $D6$ branes to place on top of the orientifolds. So it is conceivable, a priori, that what we obtain is a compactification where the $O6$ plane arrangement is different from that in the ordinary model without discrete theta angle. 

The standard reference for our current knowledge of 7d $\mathcal{N}=1$ theories is \cite{deBoer:2001wca}. Interestingly, they describe potentially not one, but \emph{two} different $O6^+/O6^-$ compactifications, which differ in the arrangement of orientifold planes. In \cite{deBoer:2001wca}, the question of whether this two components become equivalent at strong coupling was left open. From our point of view, it is natural to guess that one of them corresponds to the 7d compactification of the new component of moduli space we found in higher dimensions. This also aligns with the results of one of us in \cite{Hector-paper}, which showed there are two inequivalent embeddings of the $(D_4)^{\oplus 4}$ lattice in the K3 homology lattice; therefore, we expect two theories with these structure of frozen singularities.

It would be interesting to verify or disprove this conjecture, and check it against possible alternatives; for instance, one could also say that the T-dual of $C_2$ is naturally $\int C_3$, so the compactification corresponds to a new discrete flux on the base of $T^3$. Perhaps this description is somehow equivalent to the permuted $O6^+/O6^-$ that we described above; it would be desirable to understand this better, but we leave this task for future work. 

Instead, we spend the rest of this Section looking for additional discrete theta angles with sixteen supercharges. We begin by pursuing the M-theory approach near the end of Section \ref{sec:comp}: One can obtain new string theories in seven dimensions by compactifying M-theory on manifolds which preserve covariantly constant spinors, such as some Bieberbach manifolds. The classification of Bieberbach 4-folds is done in \cite{2013arXiv1306.6613L}, but we can get to the final answer by noting that an orientable four-manifold that preserves at least some supercharges must be hyperkahler \cite{Seiberg:1996nz}, and that the only known examples of these manifolds are $T^4$ and $K3$. Once this is established, one can look at the list of non-orientable Bieberbach manifolds constructed as circle fibrations over $T^3$; all other examples will involve as a fiber a Bieberbach other than $T^3$, and all of these do not admit covariantly constant spinors. We find that the only possibilities are
\begin{equation*}\text{M theory on}\, K3, \quad \text{KB}\times T^2,\quad \text{and}\quad N_2^3\times T^2.\end{equation*}
These are the three components of moduli space we already discussed.

It is far more productive to look at the F-theory picture instead. An F-theory background $X_d$ is a torus fibration over a base $\mathcal{B}$, where the total space $X$ is Pin$^+$ \cite{Tachikawa:2018njr}. We will look for F-theory backgrounds where the fiber does not shrink; in these cases, the $T^2$ fibration can be traded by a $\text{GL}^+(2,\mathbb{Z})$ fiber bundle over $\mathcal{B}$ \cite{Debray:2021vob}. $\text{GL}^+(2,\mathbb{Z})$ is just the duality group of IIB string theory \cite{Tachikawa:2018njr}, and what we will do here is consider compactifications with duality bundles turned on, but not 7-branes. 

Since we are compactifying to seven dimensions, we are again looking for Ricci-flat three-manifolds, which are precisely the  Bieberbach manifolds discussed in \cite{Acharya:2019mcu}.  Type IIB requires an orientation, so we restrict our attention to the orientable Bieberbach manifolds. As discussed above and in \cite{Acharya:2019mcu}, none of these admit covariantly constant spinors, except for $T^3$. But in IIB, the supercharges transform under the duality bundle, and we should look not for covariantly constant spinors, but for Spin$^{\text{GL}^+(2,\mathbb{Z})}$-covariant spinors, and several Bieberbach manifolds admit these, as we will now see. Consider a mapping torus fibration $T^2\rightarrow S^1$, where the gluing homomorphism is an element $\rho$ of $SL(2,\mathbb{Z})$. All the Bieberbach manifolds of interest (discussed below) are of this form. We can regard the resulting seven-dimensional theories as circle compactifications of type IIB on $T^2$ by an additional duality action. When going around the circle, the IIB supercharges, which transforms in a Weyl representation $\mathbb{8}$ of Spin(7,1), are transformed by an orthogonal $2\times2$ rotation matrix $M_\rho$, which is the spin lift of the isometry $\rho$. Since 8d Weyl spinors are complex, we can diagonalize to act as multiplication by phases $e^{\pm i\theta}$. This action does not leave any spinors invariant, and thus does not preserve any supersymmetry. However, we can combine it with the same action of $\rho$ embedded in the IIB duality bundle. The IIB supercharges transform as a complex spinor \cite{Pantev:2016nze,Tachikawa:2018njr} and so, for appropriate $\rho$, we can make it act as $e^{-i\theta}$. The combined action has a single surviving supercharge in 7d, leading to an $\mathcal{N}=1$ theory.

We now list all the relevant three-dimensional Bieberbach manifolds admitting covariantly constant equivariant spinors, together with the corresponding F-theory model. These three-dimensional Bieberbach manifolds all admit represesentatives as fibrations of either $T^2$ over $S^1$; each such fibration is a mapping torus associated to an element of $SL(2,\mathbb{Z})$, which implements the large diffeomorphism of the fiber as we go around the $S^1$. We now list the nontrivial Bieberbach manifolds in the notation of \cite{2013arXiv1306.6613L}, the corresponding $SL(2,\mathbb{Z})$, and the corresponding F-theory model of which the Bieberbach manifold is base:

\begin{center}
\begin{tabular}{c|c|c}
Bieberbach&$SL(2,\mathbb{Z})$ element& F-theory model\\\hline
$O^3_2$&$\begin{psmallmatrix}-1&0\\0&-1\end{psmallmatrix}$&$\frac{T^4\times S^1}{\mathbb{Z}_2}$\\
$O^3_3$&$\begin{psmallmatrix}0&-1\\1&-1\end{psmallmatrix}$&$\frac{T^4\times S^1}{\mathbb{Z}_3}$\\
$O^3_4$&$\begin{psmallmatrix}0&-1\\1&0\end{psmallmatrix}$&$\frac{T^4\times S^1}{\mathbb{Z}_4}$\\
$O^3_6$&$\begin{psmallmatrix}1&-1\\1&0\end{psmallmatrix}$&$\frac{T^4\times S^1}{\mathbb{Z}_6}$\\
\end{tabular}\end{center}

Explicitly, each of the above Bieberbach manifolds is constructed as a quotient of a parent $T^2\times S^1$ with coordinates $(\vec{x},\theta)$, each with unit period, by the isometry
\begin{equation}(\vec{x},\theta)\,\rightarrow\, (\mathbf{\rho}\cdot \vec{x},\theta+\text{ord}(\mathbf{\rho})^{-1}),\label{ebcons}\end{equation}
where $\mathbf{\rho}$ is the matrix in the second column of the table and $\text{ord}(\mathbf{\rho})$ is its order (the smallest $k$ such that $\mathbf{\rho}^k=\mathbf{I}$).

We can also compute the rank of these theories by direct dimensional reduction; this analysis will also reveal the possible discrete theta angles. Since they are all described by fibrations of $T^2$ over $S^1$, it is instructive to carry out the discussion in two steps: first form ten-dimensional IIB to eight dimensions, and then on a circle. As for the first step, we have:\begin{itemize}
\item The axio-dilaton reduces directly to an eight-dimensional complex scalar.
\item The 10d metric reduced on $T^2$ yields two KK photons, one real scalar for the volume of $T^2$, and one complex scalar for its complex structure. 
\item The $(B_2,C_2)$ fields yields the corresponding 2-forms, four vectors coming from periods on both 1-cycles of the torus,
 \begin{equation}\vec{\mathbf{A}}=\left(\int_{\text{A-cycle}} C_2,\int_{\text{A-cycle}} B_2,\int_{\text{B-cycle}} C_2,\int_{\text{B-cycle}} B_2\right),\label{rtfg}\end{equation}
as well as two axionic scalars
 \begin{equation}\vec{\phi}\equiv \left( \int_{T^2} C_2,\int_{T^2}B_2\right).\end{equation} 
 \item The $C_4$ field yields a 4-form in eight dimensions, two 3-forms, and a 2-form.\end{itemize}

 We now tackle the dimensional reduction of the 8d fields on the twisted compactification:\begin{itemize}
 \item Metric sector and axio-dilaton: Upon further reducing on the circle with a twist, the two KK photons are projected out, since all the $SL(2,\mathbb{Z})$ actions in the table above exchange the 1-cycles of the torus; one can see directly from the construction of Bieberbach manifolds as quotients  \eq{ebcons} that these isometries are not preserved. By contrast, translations along the $S^1$ base remain an isometry of the Bieberbach manifold, yielding a KK photon in seven dimensions. The volume of the $T^2$ is now accompanied by a scalar measuring the size of the $S^1$ base. The complex structure of $T^2$ and the IIB axio-dilaton are either  both frozen to special values of the moduli, or both surviving to seven dimensions, as we will explain below. The total set of seven-dimensional fields is one graviton, one vector, and either two or six real scalars.
 
\item The four-form $C_4$ and the 2-form that descends from it are insensitive to the duality bundle, producing a vector and a 2-form in seven dimensions (due to the self-duality constraint, the reduction of $C_4$ is just the magnetic potential of the circle reduction of the eight-dimensional 2-form). The three-forms coming from $C_4$ are projected out.
\item The scalars $\vec{\phi}$ above transform in the two-dimensional representation of $SL(2,\mathbb{Z})$ and so they are frozen to particular values. 
\item The vectors $\vec{A}$ transform in a four-dimensional representation of the duality and frame bundle of the $T^2$ fiber, whose details depend on $\mathbf{\rho}$. For each invariant vector, we will be able to construct a 7d vector zero mode. As we will see, we always have at least two zero modes. \end{itemize}

The minimal field content described above can be arranged into multiplets of 7d $\mathcal{N}=1$ supergravity \cite{Taylor:2011wt}. The bosonic content of the gravity multiplet consists of the graviton, 2-form (coming from $C_4$), three vectors (the KK photon, the vector coming from $C_4$, and one of the zero-modes of $\vec{A}$), and one scalar (the overall volume of the Bieberbach manifold). The other real scalar combines with  two Wilson lines to produce a vector multiplet, furnishing a 7d $\mathcal{N}=1$ theory of rank one. In those cases where the axio-dilaton and complex structure of $T^2$ are not projected out, we will find there are two additional vectors and Wilson lines, constituting two additional vector multiplets and thus enhancing the rank to three. We will now discuss each Bieberbach manifold separately, carefully analyzing the possibility of discrete theta angles:

\begin{itemize} 

\item $\mathbf{\rho}=\begin{psmallmatrix}0&-1\\1&-1\end{psmallmatrix}$: Here the axio-dilaton and torus complex structure are fixed to $e^{2\pi i/3}$, so this is a theory of rank one. The field $\vec{\phi}$ can be set to the nonzero values
\begin{equation} \vec{\phi}=\pm\left(\frac13,-\frac13\right).\end{equation}
Since multiplying these by three gives an integer, this is an example of a discrete $\mathbb{Z}_3$-valued theta angle. Both nonzero values are actually equivalent, since the $SL(2,\mathbb{Z})$ transformation that acts by multiplication by minus the identity matrix remains a valid symmetry and maps one to the other. The classification in \cite{Hector-paper} indeed predicts the existence of a discrete theta angle for F-theory on $(T^4\times S^1)/\mathbb{Z}_3$; the fact that we are able to identify discrete theta angles  precisely in the cases identified in \cite{Hector-paper} further vindicates the rules identified there.

We also need to check if there can be theta angles coming from the Wilson lines $\vec{A}$. The four-vector \eq{rtfg} has to be invariant under the combined $SL(2,\mathbb{Z})$ action on the cycles and of the duality group. In the basis specified above, this is $\mathbf{\rho}\otimes\mathbf{\rho}$, and since Wilson lines are identified up to large gauge transformations, the equation to solve to find the space of Wilson lines is
\begin{equation} \left[(\mathbf{\rho}\otimes\mathbf{\rho})-\mathbf{I}\right]\cdot\vec{A}\in\mathbb{Z}^4.\label{elret}\end{equation}
The matrix $\left[(\mathbf{\rho}\otimes\mathbf{\rho})-\mathbf{I}\right]$ has two zero eigenvalues, corresponding to the scalars of the Wilson lines in the vector multiplet described above. The question of discrete theta angles coming from Wilson lines is whether the space of solutions to \eq{elret} is connected or not. But it can be checked that the most general solution to the equation above is to take $\vec{A}$ to be a vector of integer coordinates, plus an element of the kernel. Since large gauge transformations shift $\vec{A}$ by an integer, it follows that the space is connected, and we get no discrete theta angles.

Irrespectively of the value of the discrete $\mathbb{Z}_3$ theta angle described above,  the lattice of charged states under the vectors in the gravity multiplet and the single vector multiplet does not contain a full lattice of BPS states. To see this, consider the sublattice of the charge lattice spanned by states charged under the two surviving components of $\vec{A}$. These correspond to $(p,q)$ strings wrapped on the cycles of the $T^2$ fiber. Denoting the four charges by the vector $\vec{q}=(q_1,q_2,q_3,q_4)$, so that the coupling to the vector is $\vec{q}\cdot \vec{A}$, the charges under the components invariant under the action of duality and geometry are
\begin{equation} (q_1+q_2+q_4,-q_2+q_3).\label{ab4}\end{equation}
These are simply the inner product of $\vec{q}$ by the generators of the kernel of $\mathbf{\rho}\otimes\mathbf{\rho}$. We can see that, in this normalization, these states span a charge lattice of $\mathbb{Z}^2$. Yet BPS states correspond to vectors $\vec{q}$ which are invariant under the action of the fibration on the charges, which is given by the matrix $([\mathbf{\rho}\otimes\mathbf{\rho}]^{-1})^T$. Using \eq{ab4}, the two BPS states generate the sublattice spanned by
\begin{equation} (1,1)\quad\text{and}\quad (-1,2),\end{equation}
which has index 3. 

Since the 2-form in the gravity multiplet descends from a period of $C_4$ on the $T^2$ fiber, the charged object is a $D3$ brane wrapping the $T^2$. When the discrete theta angle coming from $\vec{\phi}$ vanishes, the $D3$ wrapping $T^2$ is BPS, and we get a full lattice of BPS strings. However, things are more interesting when the theta angle is non-vanishing. It turns out that a $D3$ brane wrapped on a 2-cycle with nonvanishing periods $\int C_2,\int B_2$ has induced $(p,q)$-string charge. That this has to be the case can be easily deduced from T-duality: A $D3$ wrapped on $T^2$ an nonzero $\int C_2$ is T-dual to a $D1$ with non-zero value of $C_0$, which acquires fundamental string charge as described in Section \ref{sec:new9d}. By S-duality, we also learn that $\int B_2$ induces fundamental string charge. This picture can be made more quantitative by studying the mixed 't Hooft anomalies of the worldvolume theory of the D3 \cite{Hsieh:2019iba}. The worldvolume theory of a $D3$ is $U(1)$ $\mathcal{N}=4$ SYM, which has both electric and magnetic 1-form symmetries \cite{Gaiotto:2014kfa}. The bulk field $B_2$ acts as the background connection for the electric 1-form symmetry, and $C_2$ plays the role of the magnetic 1-form symmetry, as can be read from the two-derivative expansion of the DBI + CS action of the brane \cite{Johnson:2000ch}:
\begin{equation}S_{\text{DBI+CS}}\supset \int_{D3} \vert F-B_2\vert^2+ \int F\wedge C_2.\end{equation} 
These two 1-form worldvolume global symmetries have a mixed 't Hooft anomaly, described by the 5d auxiliary anomaly theory (see e.g. \cite{Hsieh:2019iba})
\begin{equation} \int_{M_5} B_2 \wedge dC_2.\label{ditt}\end{equation}
What this means is that the phase of the partition function of the worldvolume theory of a $D3$ is not invariant under, say, $C_2$ gauge transformations if nontrivial $B_2$ is turned on, and vice-versa. Let us consider the case of interest, where the $D3$ is wrapped on a $T^2$ with nonzero $\int B_2$. Taking $M_5=T^2\times M_3$ and reducing \eq{ditt} on $T^2$, we obtain the anomaly theory of the string  of wordlvolume $\Sigma=\partial M_3$
\begin{equation} \left(\int_{T^2} B_2\right) \int_{M_3} dC_2=\phi_1 \int_{M_3} dC_2=\phi_1 \int_{\Sigma}C_2.\end{equation}
Thus, we recover that $\phi_1$ induces $D1$ brane charge. Either explicit computation or $SL(2,\mathbb{Z})$ covariance gives the induced $(p,q)$ string charge induced by nonzero $\vec{\phi}$, 
\begin{equation} Q_{D1}=\phi_1,\quad Q_{F1}=-\phi_2.\end{equation}
Therefore, turning on the discrete $\mathbb{Z}_3$ theta angle described above means that the string of charge 1 charged under the 2-form in the gravity multiplet, obtained from a single $D3$ wrapping the $T^2$, is charged under the 2-form fields $C_2, B_2$, whose zero modes are projected out by the $SL(2,\mathbb{Z})$ action in the fiber. What this means is that the $D3$-brane sources  discrete fields, and cannot be BPS.  Again, we find that turning on the discrete theta angle affects BPS completeness of the lattice of strings in a theory with sixteen supercharges. When one considers a state with a $D3$-brane charge multiple of 3, the induced $(p,q)$ string charge is integer, and may be canceled by adding a $(p,q)$-string on top of the $D3$ brane; the system will then relax to a BPS configuration with no induced $(p,q)$-string charge. 

\item $\mathbf{\rho}=\begin{psmallmatrix}0&-1\\1&0\end{psmallmatrix}$: The analysis here is very similar to that of the previous point, so we will be brief. This is too a theory of rank 1, since $\tau$ is fixed to $i$, and $\vec{\phi}$ can be set to the nonzero value
\begin{equation} \vec{\phi}=\left(\frac12,\frac12\right).\end{equation}
There is, therefore, a discrete $\mathbb{Z}_2$ angle in this case; just as in the case above, this lives in a new component of moduli space in seven dimensions. Again, this is in agreement with the results of \cite{Hector-paper}. Turning on the discrete theta angle means that the string of elementary charge under the 2-form of the gravity multiplet is not BPS; this time, the sublattice of BPS strings has index two.  

An analysis of the Wilson lines shows there are no discrete theta angles associated to them either. The sublattice of BPS states is generated by the sublattice spanned by $(2,0)$ and $(0,2)$ in $\mathbb{Z}^2$, and therefore has index 4 (but its coarseness, as defined in \cite{Heidenreich:2016aqi}, is only 2). 

\item $\mathbf{\rho}=\begin{psmallmatrix}1&-1\\1&0\end{psmallmatrix}$: In this case, there are no discrete theta angles coming either from $\vec{\phi}$ or $\vec{A}$. We just have a single component of moduli space, there is BPS string completeness, and the lattice of BPS charged particles is of index 3. 

\item $\mathbf{\rho}=\begin{psmallmatrix}-1&0\\0&-1\end{psmallmatrix}$: We save the most complex case for last. This element of $SL(2,\mathbb{Z})$ acts trivially on the upper half plane, and so it corresponds to the case where the axio-dilaton and complex structure of $T^2$ are not projected out. The action on $\vec{A}$ is trivial, since
\begin{equation} \left(\begin{array}{cc}-1&0\\0&-1\end{array}\right)\otimes  \left(\begin{array}{cc}-1&0\\0&-1\end{array}\right)= \left(\begin{array}{cccc}1&0&0&0\\0&1&0&0\\0&0&1&0\\0&0&0&1\end{array}\right),\end{equation}
where the first tensor factor represents the geometric action on the cycles of the torus and the second one refers to the duality bundle. The resulting theory is of rank 3, as opposed to all previous examples which are of rank one, and it lives on a different component of moduli space of the usual $O6^+/O6^-$ orientifold compactification, which is dual to M-theory on KB$\times T^2$ \cite{deBoer:2001wca}.
A priori, this component of the moduli space admits two discrete theta angles, since one can set $\vec{\phi}$ equal to $(0,0)$, $(1/2,0)$, $(0,1/2)$, or $(1/2, 1/2)$. All non-zero values of these theta angles lead, by the same arguments as above, to a lattice of BPS strings of index 2. 

\end{itemize}

\begin{table}[!hbt]
\begin{center}
\begin{tabular}{c|c|c|c}
Rank&Description&$\theta$ angles& BPS strings sublattice index\\\hline
\multirow{6}{*}{3}& M on KB$\times T^2$& No&1  \\
&\cellcolor{light-blue} M on $N_2^3\times S^1$& \cellcolor{light-blue}No&\cellcolor{light-blue}2 \\
&\multirow{4}{*}{IIB on $O^3_2$}& No&1\\
&&\cellcolor{light-blue}$\int B_2=1/2$&\cellcolor{light-blue}2\\
&&\cellcolor{light-blue}$\cellcolor{light-blue}\int C_2=1/2$&\cellcolor{light-blue}2\\
&&\cellcolor{light-blue}$\int B_2=1/2,\int C_2=1/2$&\cellcolor{light-blue}2\\\hline
\multirow{5}{*}{1}&\multirow{2}{*}{IIB on $O^3_3$}& No&1\\&&\cellcolor{light-blue}$\int C_2=\frac13,\int B_2=-\frac13$&\cellcolor{light-blue}3\\
&\multirow{2}{*}{IIB on $O^3_4$}& No&1\\&&\cellcolor{light-blue}$\int C_2=\frac12,\int B_2=\frac12$&\cellcolor{light-blue}2\\
&\multirow{1}{*}{IIB on $O^3_6$}& No&1
\end{tabular}
\end{center}

\caption{Table of the low-rank ($r\leq 3$) components of the moduli space with sixteen supercharges in seven dimensions, their descriptions in terms of smooth supergravity backgrounds, including possible discrete theta angles and the index of the sublattice of BPS strings in each case. The entries shaded in blue represent the new components of moduli space discussed in this paper. In the case of the rank 3 theories, there may be dualities relating different components of moduli space, so that some of the possibilities may be equivalent. }
\label{t1}
\end{table}

The list above just describes the models and theta angles which have a description in terms of type IIB on orientable Bieberbach manifolds. To these models, we must add the dimensional reduction of M-theory on KB and the new component we found in nine dimensions, which correspond to circle fibrations with base the trivial Bieberbach manifold $T^3$. The results of this Section are summarized in Table \ref{t1}. In the table, we have also included comparison with 7d theories of sixteen supercharges discussed previously in the literature, most notably in \cite{deBoer:2001wca}. We believe the type IIB constructions in terms of Bieberbach manifolds described above capture a previously unexplored corner of the same components of the moduli space of the rank 1 theories constructed in \cite{Aharony:2007du} in terms of F/M theory with frozen singularities; the two descriptions are likely related by $T$-duality, which can introduce singularities (as in the AOB background being dual to an $O7^+/O7^-$ compactification). Furthermore, the charge lattices of both theories match. It would be interesting to check this conjecture further.

The situation is particularly interesting for the theories of rank 3. In \cite{deBoer:2001wca}, these theories were described via $O6^+/O6^-$ compactifications; it was noted in there that there is more than one possible inequivalent arrangement of the $O6^+/O6^-$  planes after diffeomorphisms are taken into account, suggesting the presence of at least two components of moduli space at rank 3. Reference \cite{deBoer:2001wca} then speculated that these components might be equivalent at the non-perturbative level, if the corresponding embeddings of their charge lattice into the K3 charge lattice turned out to be equivalent.  The results we have obtained here suggest instead that these two components of moduli space are inequivalent. In fact, we find what looks like \emph{five} distinct components of moduli space of rank 3: two of them descend from the eight dimensional components obtained from $KB\times S^1$ and $N_2^3$,  and the last three come from the last entry in the table above with different nonzero choices of discrete theta angles (with theta angle turned off, we believe this is dual to the component on $KB\times T^2$). It would be an interesting question to elucidate the structure of this component of moduli space, and whether some of these theories are dual to each other, or not, but we will not try to solve this here. 

In the process of working out the supergravity description of these theories, we uncovered two new discrete theta angles, producing two new components of moduli space, as noted in the Table. This exactly matches the predictions by one of us in \cite{Hector-paper}, based on the fact that there are two inequivalent ways to freeze the corresponding singularities in $K3$. An outstanding question is the description of these theta angles in the M/F theory picture; the inequivalent lattice embeddings seems to suggest the tantalizing possibility that in F-theory there is also the notion of ``freezing a section'' of the elliptic fibration, and not just a singularity. Formalizing our rudimentary understanding of this phenomenon, and its extension to lower supersymmetry, is a very interesting question we hope to come to in the future. 

All in all, we recover the theories described in the last five entries of Table 1 of \cite{deBoer:2001wca}, together with the F-theory description of our new component in moduli space, but no new theories. There is one more orientable Bieberbach manifold, of holonomy $\mathbb{Z}_2\times\mathbb{Z}_2$, that does not preserve any supersymmetry. The Bieberbach descriptions exhibit these theories as completely smooth type IIB compactifications, and we can now look for discrete theta angles as in the rest of this paper. $C_0$ is already accounted for by our previous discussion, so the only possible holonomies are that of the $(B_2,C_2)$ fields. 

\section{Discrete theta angles in non-supersymmetric string theories}\label{sec:nonsusy}
The main theme of this paper is to establish that discrete theta angles, far from being exotic, are a very common feature of string compactifications, and can lead to different physics even for highly non-supersymmetric theories. We will now study the possibility of discrete theta angles in the existing ten-dimensional non-supersymmetric string theories in ten dimensions, to see if we can construct any new examples. We only know three examples of non-supersymmetric string theories in ten dimensions, so enlarging this landscape could be significant.

We will first discuss discrete theta angles in the non-supersymmetric Sugimoto string \cite{Sugimoto:1999tx}. The Sugimoto theory is a nonsupersymmetric variant of the construction of type I string theory. The latter is constructed as an orientifold of IIB in ten dimensions, with an $O9^-$ plane and $32$ $D9$ branes to cancel the tadpole. The Sugimoto string is constructed by instead replacing the $O9^-$ by an $O9^+$. This has the opposite RR charge than the $O9^-$, and to cancel the tadpole, one must introduce 32 \emph{anti}-$D9$ branes. This makes the resulting background nonsupersymmetric, but still amenable to a worldsheet description.  Unlike type I, the Sugimoto string does not admit a discrete RR theta angle. One way to see this is that Sugimoto has symplectic gauge groups, and 
\begin{equation} \pi_9(\text{Sp}(16))=0.\end{equation}
Relatedly, the tachyon in the $D(-1)$-$\overline{D(-1)}$ IIB system, which is projected out by the ordinary type I projection, remains in the Sugimoto model \cite{Witten:1998cd}. There is no stable charged $\mathbb{Z}_2$ instanton.

We now pass to the $SO(16)\times SO(16)$ heterotic string \cite{Alvarez-Gaume:1986ghj}. Since
\begin{equation} \pi_9(SO(16)\times SO(16))=\mathbb{Z}_2\oplus\mathbb{Z}_2,\end{equation}
there is room for two different discrete theta angles. At the massless level, the theory contains fields transforming in one of the spinorial representations of each $SO(16)$ factor, and this strongly breaks any would-be $O(16)\times O(16)$ symmetry \cite{Alvarez-Gaume:1986ghj}, since these transformations would reverse the chirality of the spinors but not the vectors, spoiling local anomaly cancellation. One then could conclude that these theta angles seem quite physical, producing a total of three new non-supersymmetric ``cousins'' of the $SO(16)\times SO(16)$ string. It would be very interesting to explore whether these angles are really there and if so, what are their physical effects.

Lastly, there is a third non-supersymmetric model, with gauge group $U(32)$ \cite{Sagnotti:1996qj}, obtained an orientifold of the non-supersymmetric 0B string \cite{Blumenhagen:1999bd}. The model admits a continuous theta angle, which couples to one of the two RR axions of type 0B, but no obvious discrete theta angles.

\section{Conclusions}\label{sec:conclus}
We know remarkably little about the properties of quantum gravity vacua away from the low-energy supergravity regime. We can sometimes get a handle on this via discrete theta angles, topological couplings that are invisible at the massless level but which can be studied reliably. In this note, we have studied theta angles in string compactifications with sixteen supercharges to produce several new string theories, with and without supersymmetry, in nine, eight, and seven dimensions.

Our work clarifies in which circumstances discrete theta angles arising in string compactifications actually exist. Sometimes they may seem to be there, but are unphysical, since there are no charged instantons that can possibly detect them. This is the case of the Sethi string \cite{Sethi:2013hra}, an intriguing proposal for a new supersymmetric string theory in ten dimensions. We have now shown it is actually exactly equivalent to ordinary type I string theory. Although the particular construction in \cite{Sethi:2013hra} fails, the general idea does work; and this is how we have uncovered new string theories in this paper. 

In nine dimensions, we have discovered a new connected component of moduli space, separate from the two already known ones, and related to one of them by a discrete theta angle. The most outstanding feature of this theory, apart from enlarging the Landscape of nine-dimensional compactifications, is that it violates the BPS completeness principle for string charges \cite{Kim:2019vuc}, the idea that with enough supersymmetry there must be BPS states for all quantized values of the charge. For some charges, in our examples, we only find non-BPS representatives, in agreement with the Completeness Principle \cite{Polchinski:2003bq,Banks:2010zn} but not with its BPS extension. Although counterexamples to this statement were known in the case of particles and with lower supersymmetry \cite{Heidenreich:2016aqi}, all previously known examples with sixteen supercharges had a full lattice of BPS strings. In the new theory we have found, the BPS states form a sublattice of index two.

The violation of BPS completeness has important consequences for the Swampland, since several papers (e.g. \cite{Kim:2019vuc,Lee:2019skh,Lanza:2019xxg,Lanza:2020qmt,Kim:2019ths,Katz:2020ewz,Angelantonj:2020pyr,Cvetic:2020kuw,Tarazi:2021duw,Cvetic:2021vsw}) use BPS completeness to rule out certain seemingly consistent supergravities. Based on the above, the conclusions of these papers should be read instead as showing the impossibility of having both these theories and a complete spectrum of BPS strings. 

While BPS completeness fails at the pure supergravity level, taking into account discrete symmetries present in the compactification allows a microscopic understanding of why there are no BPS representatives of every charge. We believe likely that a suitable generalization of BPS completeness, including the effects of additional discrete symmetries, does hold and captures these examples as well. What is clear however is that such a notion goes beyond the pure supergravity regime.

Compactifying this theory on a circle allows one to access a dual M theory description in eight dimensions. The theory is simply M-theory compactified on a certain fibration of the Klein bottle on a circle, which happens to preserve supersymmetry.  This background, while non-orientable, is a perfectly smooth and standard background of M-theory. It would be interesting to study this background in other corners of the duality web and find non-perturbative information about the spectrum of massive states, perhaps along the lines of \cite{Cvetic:2021sjm,Cvetic:2022uuu}.

\begin{figure}[hbtp!]
\centering 
    \includegraphics[width=0.85\textwidth]{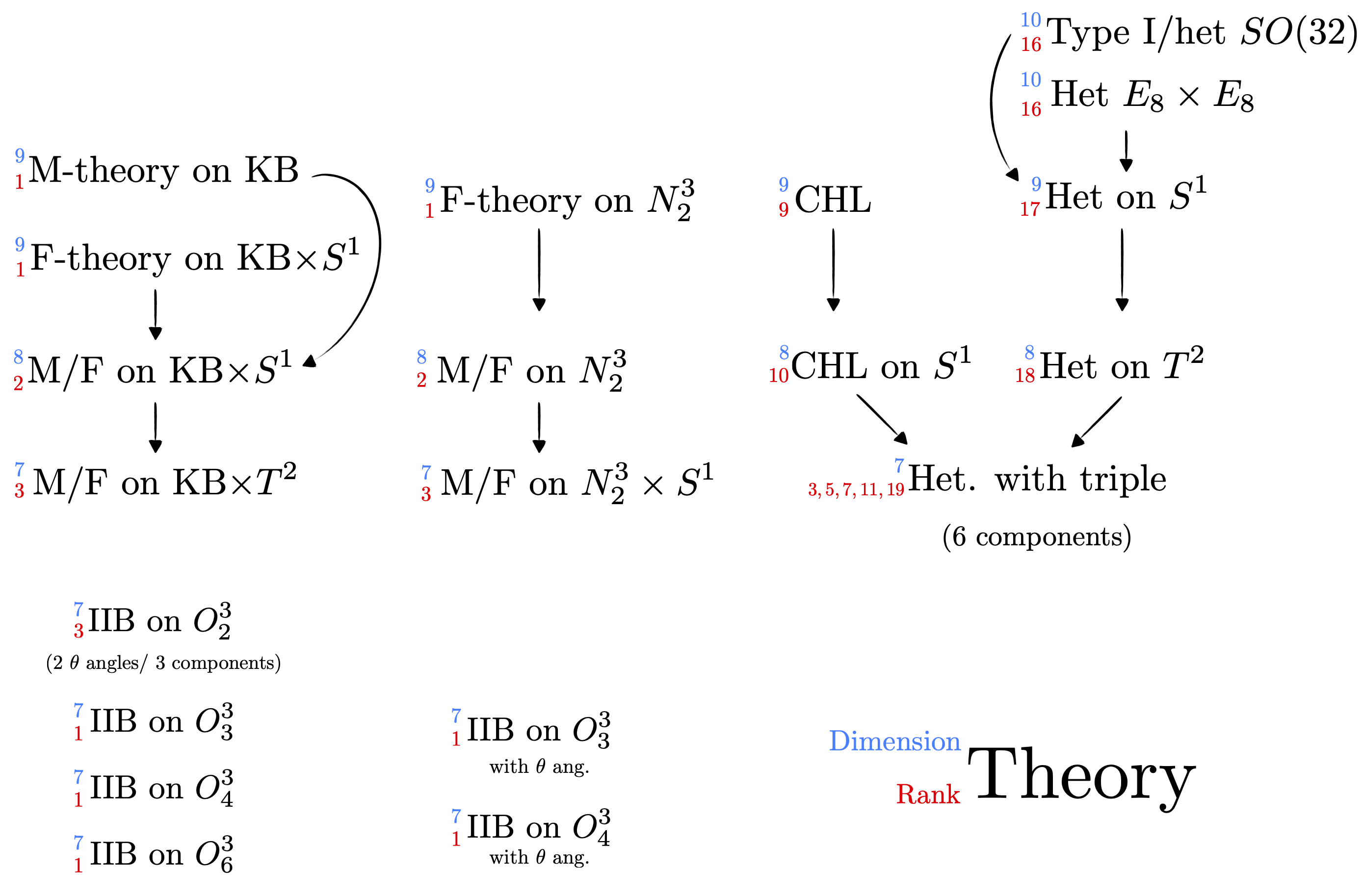}
    \caption{List of all known disconnected theories with 16 supercharges in dimension $\geq 7$. We label each component of moduli space by a representative compactification (e.g. heterotic on $S^1$); we do not list all dual descriptions. Theories connected by an arrow are related by dimensional reduction. For each component we indicate both the dimension (upper left blue number) and the rank (lower left red number), in a manner meant to be reminiscent of the periodic table of elements. }
    \label{fsumm}
\end{figure}

In seven-dimensional theories we proposed a new description of the lower-dimensional rank theories, in terms of IIB compactified on Bieberbach manifolds. While these Riemann-flat manifolds do not admit covariantly constant ordinary spinors, they do admit covariantly constant $SL(2,\mathbb{Z})$ spinors, and so they can yield supersymmetric backgrounds in IIB. This approach provides a new description of part of the moduli space of these theories, and allowed us to discover two new components associated to discrete theta angles coming from 2-form fields. Turning on these theta angles has the effect of increasing the index of the sublattice.  A summary of the known components of moduli space with sixteen supercharges and more than six dimensions can be found in Figure \ref{fsumm}.

Both the new theory we found in 9d as well as the new description of the 7d theories come from compactifying the familiar ten-dimensional string theories or M-theory in Bieberbach manifolds, quotients of tori that constitute the most general Riemann-flat manifolds. The classification of all possible supersymmetric compactifications on Bieberbach manifolds has not been worked out, and can potentially help us uncover even more new components of moduli space, and is a problem to which we hope to return in the near future. 

Interestingly, with the description of the 7d theories in terms of smooth Bieberbach manifolds, it is now true that all theories with 16 supercharges admit a region of their moduli space which is well approximated by smooth compactification of a ten or eleven-dimensional supergravity, without adding singular brane sources. This means that, at least in $d\geq7$, supergravity techniques are enough to fully capture all components of moduli space. It would be interesting to find out if this feature persist in lower dimensions or with fewer supercharges\footnote{See \cite{Cheng:2022nso} for recent work involving Bieberbach manifolds in regimes with eight supercharges. }, potentially leading to the conjecture that all components of moduli space in quantum gravity can be detected by supergravity, without needing to resort to stringy techniques. 

We have also explored the question of discrete theta angles in the ten-dimensional nonsupersymmetric string theories, potentially finding examples in the $SO(16)\times SO(16)$ theory, but none in the Sugimoto or 0B' strings. It would be interesting to explore this in more detail in the future, and determine whether the $SO(16)$ theta angles are physical or not.

Our results illustrate dramatically that one can have several different quantum gravities, which are identical at the massless level, but have radically different sets of massive states. This has a large impact on foundational questions within the Swampland program, such as what is the minimal charge which satisfies the Weak Gravity Conjecture.  At least in some examples, these differences can be captured by topological couplings, and thus while formally out of reach of supergravity, a classification and rigorous analysis are possible. All these examples yield new compactifications, and new potential avenues for string model building. We believe that by exploring more examples with discrete theta angles we enlarge not only our knowledge of the extension of the Landscape, but also of the limits of the Swampland.

\vspace{0.5cm}

\textbf{Acknowledgements}: We thank Oren Bergman, Markus Dierigl, I. Garc\'{i}a-Etxebarria, Yuta Hamada, Jonathan Heckman, Ben Heidenreich, Simeon Hellerman, Jake McNamara, Paul Oehlman, Cumrun Vafa, Irene Valenzuela, and Alberto Zaffaroni for many useful discussions and comments on the draft. MM is grateful to the Simons Center Summer Workshop 2022 for hospitality and a stimulating environment while this work was completed, as well as to the ``Engineering in the Landscape'' workshop at Uppsala University.  HPF is grateful for the hospitality  of the CMSA at Harvard University, under the CMSA Swampland Program.  The work of MM is supported by a grant from the Simons Foundation (602883, CV). This work was also partially supported by the ERC Consolidator Grant 772408-Stringlandscape.

\appendix

\section{Elements of \texorpdfstring{$GL(2,\mathbb{Z})$}{GL(2,Z)} with fixed points}\label{appA}
In Section \ref{subsec:dual} we described a few elements of $GL(2,\mathbb{Z})$ of determinant $-1$ and fixed points, corresponding to the AOB and DP backgrounds, as well as their versions with theta angles turned on. We perform here a more general analysis. We consider a general element $M\in GL(2,\mathbb{Z})$ of determinant $-1$, described as a matrix
\begin{equation}M=\begin{pmatrix}a&b\\c&d\end{pmatrix},\quad ad-bc=-1\end{equation}
and copy the fixed point equation \eq{bilet} from the main text,
\begin{equation}
	c y^2 + c x^2  -(a-d)\, x - b = 0 \,, ~~~~~ (a+d)\,y = 0\,.
\end{equation}
Since we are looking for solutions in the upper half plane, $y>0$, the second equation implies $a=-d$. Since the first equation is symmetric under $y \mapsto -y$, restricting to positive $y$ is always possible. To find the most general solution, there are three separate cases to consider:
\begin{enumerate}[i)] 

\item Setting $a = 0$, we have $cy^2 + cx^2 - b = 0$ with $c \neq 0$ (this would give $M = 0$), so that
\begin{equation}
	x^2	+ y^2 = \frac{b}{c}\,, ~~~~~ (a = 0)\,,
\end{equation}
But on the other hand, $bc = 1$, and so 
\begin{equation}
	M = \pm\begin{pmatrix}
	0 & 1 \\ 1 & 0
	\end{pmatrix}\,, ~~~~~ (a = 0)\,.
\end{equation}
\item Setting $c = 0$, we have that $2ax - b = 0$ with $ad-bc = -a^2 = -1 \Rightarrow a = \pm 1$ hence
\begin{equation}
	\pm 2x - b = 0\,.
\end{equation}
The corresponding family of solutions is
\begin{equation}
	M(b) = \pm \begin{pmatrix}1 & b \\ 0 & -1\end{pmatrix}\,,
\end{equation}
which fix $\tau = -\frac{b}{2} + iy$.

\item Setting $b = 0$, we have $cy^2+cx^2-2ax = 0$, again with $a = \pm 1$, so that
\begin{equation}
	x^2 + y^2 \pm \frac{2}{c}x = 0\,.
\end{equation}
This gives the family of solutions
\begin{equation}
	M(c) = \pm \begin{pmatrix}1 & 0 \\ c & -1	\end{pmatrix}\,,
\end{equation}
which fix $\tau = x + i\sqrt{x(2/c - x)}$ with $0<x<2/c$ for positive $c$ and $c/2>x>0$ for negative $c$.

\item The general case is, in the chart $c \neq 0$, given by
\begin{equation}
	M = \begin{pmatrix} a & \frac{1-a^2}{c} \\ c & -a	\end{pmatrix}\,, ~~~~~ \tau_\text{fixed} = x + i\sqrt{-x^2 + \frac{2a}{c}x + \frac{1-a^2}{c^2}}\,.\label{e2ff3}
\end{equation}
Notice that, for this to be an element of $GL(2,\mathbb{Z})$, $c$ should divide $1-a^2$. The elements in this class give all the images of the curve $\vert\tau\vert^2 = 1$ under $GL(2,\mathbb{Z})$ except for those with $c = 0$, namely vertical lines. This suggests that the elements in this class are precisely $GL(2,\mathbb{Z})$ images of the members in the above two classes, which as shown in the main text, correspond to the class of the AOB and DP backgrounds. Although we have not proved this, we have verified that the image of ii) and iii) under a general $SL(2,\mathbb{Z})$ element is generically of the form \eq{e2ff3}. If this is the case, all the additional $GL(2,\mathbb{Z})$ conjugacy classes correspond to (in general intrinsically coupled descriptions) of the AOB and DP backgrounds, with and without theta angle, and hence do not provide new theories in nine dimensions.
\end{enumerate}

\bibliographystyle{JHEP}
\bibliography{discrete-theta-refs}

\end{document}